\documentclass[fleqn,usenatbib]{mnras}

\usepackage{mathptmx}
\usepackage[T1]{fontenc}
\usepackage{ae,aecompl}

\usepackage{graphicx}	% Including figure files
\usepackage{amsmath}	% Advanced maths commands
\usepackage{amssymb}	% Extra maths symbols
\usepackage{xspace}
\usepackage[dvipsnames]{xcolor}

\newcommand{\mufasa}{{\sc Mufasa}\xspace}
\newcommand{\simba}{{\sc Simba}\xspace}

\newcommand{\lya}{Ly$\alpha$\xspace}
\newcommand{\HI}{\ion{H}{i}\xspace}
\newcommand{\MgII}{\ion{Mg}{ii}\xspace}
\newcommand{\SiIII}{\ion{Si}{iii}\xspace}
\newcommand{\CIV}{\ion{C}{iv}\xspace}
\newcommand{\OVI}{\ion{O}{vi}\xspace}

\newcommand{\hkpc}{h^{-1}{\rm kpc}}

\newcommand{\hmpc}{h^{-1}{\rm Mpc}}

\newcommand{\ghi}{\Gamma_{\rm HI}}
\newcommand{\kms}{\;{\rm km}\,{\rm s}^{-1}}

\newcommand{\msolar}{\;{\rm M}_{\odot}}

\newcommand{\gizmo}{{\sc Gizmo}\xspace}
\newcommand{\caesar}{{\sc Caesar}\xspace}
\newcommand{\pyigm}{{\sc PyIGM}\xspace}

\newcommand{\fedd}{f_{\rm Edd}}
\newcommand{\mbh}{M_{\rm BH}}

\newcommand{\mstar}{\mbox{$M_\star$}\xspace}
\newcommand{\mhalo}{\mbox{$M_{\rm halo}$}\xspace}
\newcommand{\pygad}{{\sc Pygad}\xspace}
\newcommand{\music}{{\sc Music}\xspace}
\newcommand{\cloudy}{{\sc Cloudy}\xspace}
\newcommand{\grackle}{{\sc Grackle}\xspace}

\newcommand{\fcov}{$f_{\rm cov}$\xspace}
\newcommand{\lstar}{L$^\star$\xspace}

\definecolor{mycolor}{rgb}{255,0,0}
\def\mycolor{\textcolor{black}}
\def\Romeel{\textcolor{black}}
\def\response{\textcolor{black}}

\title[CGM in Simba]{The Low Redshift Circumgalactic Medium in \simba}

\author[S. Appleby et al.]{
Sarah Appleby$^{1}$\thanks{E-mail: sapple@roe.ac.uk},
Romeel Dav\'e$^{1,2,3}$,
Daniele Sorini$^{1}$, 
Kate Storey-Fisher$^{4}$
\newauthor \& Britton Smith$^{1}$
\\
\\$^{1}$ SUPA\thanks{Scottish Universities Physics Alliance}, Institute for Astronomy, University of Edinburgh, Royal Observatory, Edinburgh EH9 3HJ, UK
\\$^2$ University of the Western Cape, Bellville, Cape Town 7535, South Africa
\\$^3$ South African Astronomical Observatories, Observatory, Cape Town 7925, South Africa
\\$^4$ Center for Cosmology and Particle Physics, Department of Physics, New York University,
New York, NY, USA
}

\date{Accepted XXX. Received YYY; in original form ZZZ}

\pubyear{2021}

\begin{document}
\defcitealias{haardt_2001}{HM01}
\defcitealias{haardt_2012}{HM12x2}
\defcitealias{faucher-giguere_2020}{FG20}
\label{firstpage}
\pagerange{\pageref{firstpage}--\pageref{lastpage}}
\maketitle

\begin{abstract}
We examine the properties of the low-redshift circumgalactic medium (CGM) around star-forming and quenched galaxies in the \simba cosmological hydrodynamic simulations, focusing on comparing \HI and metal line absorption to observations from the COS-Halos and COS-Dwarfs surveys. 
Halo baryon fractions are \Romeel{generally $\la 50\%$ of} the cosmic fraction due to stellar feedback at low masses, and jet-mode AGN feedback at high masses.
Baryons and metals in the CGM of quenched galaxies are \Romeel{ $\ga 90\%$} hot gas, while the CGM of star-forming galaxies is more multi-phase.
Hot CGM gas has low metallicity, while warm and cool CGM gas have metallicity close to that of galactic gas.
Equivalent widths, covering fractions and total path absorption of \HI and selected metal lines (\MgII, \SiIII, \CIV and \OVI) around a matched sample of \simba star-forming galaxies are mostly consistent with COS-Halos and COS-Dwarfs observations \Romeel{to $\la 0.4$~dex, depending on ion and assumed ionising background.}
Around matched quenched galaxies, absorption in all ions is lower, with \HI absorption \Romeel{significantly} under-predicted.
Metal-line absorption is sensitive to choice of photo-ionising background; assuming recent backgrounds, \simba matches \OVI but under-predicts low ions, while an older background matches low ions but under-predicts \OVI.
\simba reproduces the observed dichotomy of \OVI absorption around star forming and quenched galaxies.
CGM metals primarily come from stellar feedback, while jet-mode AGN feedback reduces absorption particularly for lower ions.
\end{abstract}


%%%%%%%%%%%%%%%%%%%%%%%%%%%%%%%%%%%%%%%%%%%%%%%%%%

%%%%%%%%%%%%%%%%% BODY OF PAPER %%%%%%%%%%%%%%%%%%

\section{Introduction}

% Why do we care about the CGM?
Galaxies interact dynamically with their environment. They accrete new material for star formation, either pristine gas from the intergalactic medium \citep[IGM;][]{keres_2005,dekel_2009}, or recycled gas that has been processed through star formation \citep{oppenheimer_2010, van_de_voort_2011}.  Galaxies also eject material via energetic outflows from supernova events and/or black hole activity~\citep{Veilleux_2005}. Additionally, they can lose material via mergers or due to environmental processes such as ram pressure stripping.  These processes, collectively referred to as the baryon cycle, are believed to be a primary regulator of how galaxies form and evolve.  Any material entering or exiting a galaxy halo must pass through the circumgalactic medium (CGM). Therefore the CGM is a potentially rich source of information about galaxy evolution.

% How do we normally measure the CGM?
In the local universe, the CGM of star-forming galaxies is typically probed via absorption lines in the spectra of background quasars passing  at small impact parameter from the foreground galaxies. Such observations show that the CGM is a multiphase, complex environment with many physical origins \citep{chen_2017, tumlinson_2017}.  More massive quenched galaxies are often associated with X-ray halos~\citep{mulchaey_2000,rosati_2002}, showing that their CGM is predominant in the form of hot gas.  Nonetheless, absorption line observations around massive galaxies show the presence of cool gas \citep{thom_2012,zahedy_2019}, indicating that the CGM is multi-phase at all halo mass scales. Given that absorption lines can typically probe to much lower gas densities than emission measures such as X-ray and resonant line maps using integral field units, they hold great promise in being able to elucidate the physical and dynamical conditions in the CGM across a wide range of galaxy masses and types.

The primary drawback of absorption-line measurements is that typically each galaxy has only one or at most a few lines of sight by which to extract information. Current absorption line samples also remain relatively small because many of the key transitions are in the ultraviolet, requiring space-based observations at low redshifts or else optical observations of fainter distant sources at high redshifts.  Unfortunately, the intrinsic variations in the CGM can be large even within a narrow range of galaxy or halo masses~\citep{hani_2019}, which means that fully characterising the diversity of CGM in absorption is observationally daunting.

One way to augment such observational studies towards a more physical picture is to look at the CGM in cosmological hydrodynamic simulations \citep{ford_2013, hummels_2013, ford_2014, ford_2016, nelson_2018b, oppenheimer_2018b, hafen_2019, van_de_voort_2019, peeples_2019, hafen_2020}.  If such simulations reasonably reproduce the observed absorption line statistics and other galaxy properties, they could potentially offer a full 3-D view of the CGM, and enable us to interpret the physical and dynamical state of the absorbing gas, at least within a context of a given model.  Moreover, simulations used as numerical experiments allow us to test the relationship between CGM observables and galactic feedback processes, in particular star formation feedback at lower mass scales and AGN feedback in $\ga L^\star$ galaxies.  As such, simulations have become an important tool in understanding the baryon cycle via CGM absorption line data~\citep{tumlinson_2017}. 

At low redshifts, {\it Hubble's} Cosmic Origins Spectrograph (COS) provides the main datasets for constraining models.  Among key ideas emerging from low-$z$ CGM simulations are that low ions arise in cool, dense gas located closer to galaxies, while high ions like \OVI arise more in extended hotter gas that can be heated from virialisation as well as feedback~\citep{ford_2013,hummels_2013}.  \citet{ford_2014} used particle tracking to show that high ions trace metals deposited long ago via outflows, while low ions trace inflowing/outflowing gas within the past few Gyr.  Comparing these same simulations to the COS-Halos sample, \citet{ford_2016} found good agreement for \HI and some metal lines and an underproduction of \SiIII and \OVI. \citet{gutcke_2017} used \HI and \OVI absorption to trace cold and hot CGM gas in the NIHAO simulation suite \citep{wang_2015}, finding that the inner CGM is dominated by extended gas disks. The EAGLE simulation \citep{crain_2015, schaye_2015} reproduces COS-Halos observations of low metal ion absorption, arising from cool clouds in a hot ambient medium \citep{oppenheimer_2018b}. When including time-variable AGN radiation in EAGLE to mimic overionisation in the CGM, \OVI absorption can increase to the level needed to match COS-Halos \citep{oppenheimer_2018a}. IllustrisTNG \citep{pillepich_2018} predicts \OVI absorption in good agreement with COS-Halos and reproduces the dichotomy of \OVI absorption seen between star forming and quenched galaxies \citep{nelson_2018b}. In the Romulus25 simulations \citep{tremmel_2017} \OVI absorption traces black hole mass and accretion history \citep{sanchez_2019}.  \Romeel{These examples illustrate how comparisons to low-$z$ CGM absorbers, primarily from COS, provide valuable insights on the physical state of CGM gas along with constraints on galaxy formation simulations.}

Here we use the \simba\ cosmological hydrodynamic simulation to explore the CGM around low-redshift galaxies. \simba\ has been shown to reproduce a wide variety of galaxy observations at low redshifts~\citep[e.g.][]{dave_2019,li_2019,thomas_2019,dave_2020,appleby_2020,thomas_2021}, as well as low-redshift \lya\ statistics along random lines of sight~\citep{christiansen_2020} particularly when employing the recent \citet{faucher-giguere_2020} model of the ionising background.  Being a cosmological simulation, \simba\ has $\sim$kpc resolution within dense regions, degrading in the CGM owing to the adaptive gas smoothing scale.  This means it does not have as high resolution as zoom simulations, particularly ones that target the CGM, which is likely to most impact the statistics of low-ionisation absorbers that come from the densest CGM gas~\citep{van_de_voort_2019,peeples_2019} while having lower impact on mid- and high-ionisation absorbers that are most commonly detected in the CGM.  On the other hand, a major advantage is that \simba\ models a representative cosmological volume of 100 $\hmpc$, enabling statistical comparisons against observations, including being able to mimic selection criteria on host galaxy properties, while at the same time consistently reproducing IGM statistics \citep[see e.g.][]{sorini_2020}.  

In this work, we compare \simba\ CGM absorption directly to the COS-Halos~\citep{tumlinson_2013} and COS-Dwarfs~\citep{bordoloi_2014} samples.  These samples span three orders of magnitude in stellar mass, including both star-forming and quenched systems.  By performing a careful comparison to these data using mock absorption spectra around similarly-selected galaxies, we can examine the CGM around a wide range of galaxy types, providing a detailed test of \simba's CGM predictions as well as insights into the physical state of the CGM gas.  Moreover, by using \simba\ variants with individual feedback modules turned on and off, we can perform numerical experiments to understand the key physical drivers of CGM absorption within \simba, as well as test the impact of the assumed photoionising background among various recent determinations thereof.  This work builds and extends that of \citet{ford_2016} in three main ways: (i) we study the CGM around star-forming and quenched galaxies separately, enabled by the observationally-concordant dichotomy in the galaxy population from \simba; (ii) we study the correlation between input physics and CGM properties using \simba's variant runs with individual feedback modules turned on and off; and (iii) we examine the effect of assuming several different meta-galactic photo-ionising backgrounds.
%In general, we find that \simba\ broadly reproduces the properties of CGM absorbers around both star-forming and quenched systems, in terms of amplitude and trends versus radius.  Inclusion of AGN feedback makes the most difference for the hot CGM around quenched galaxies, and varying the assumed ionising background between \citet{haardt_2001} and \citet{faucher-giguere_2020} or \citet{haardt_2012} makes a substantial difference particularly for \ion{Si}{iii} and \ion{O}{vi}, the latter relieving some tensions found in earlier works~\citep{ford_2013,hummels_2013}.

% Outline the paper
This paper is organised as follows.  In \S\ref{sec:sims} we present the
\simba simulations. In \S\ref{sec:mass_metal_budget} we present the mass and metal budgets of the $z=0$ galaxy halos in \simba. In \S\ref{sec:cos_comparison} we compare \HI and metal line absorption in \simba to the COS-Halos and COS-Dwarfs surveys. In \S\ref{sec:feedback} we examine the effect of stellar and AGN feedback on the CGM. Finally in \S\ref{sec:conclusions} we conclude and summarise. Unless explicitly stated otherwise, distance units are expressed in co-moving coordinates.

\section{Simulations}\label{sec:sims}

% Big picture stuff
We use the \simba simulation \citep{dave_2019}, which is the successor to the \mufasa simulation \citep{dave_2016} and was run using a modified version of the gravity plus hydrodynamics code \gizmo \citep{hopkins_2015} in its Meshless Finite-Mass (MFM) mode. The fiducial \simba simulation models a 100 $\hmpc$ comoving volume from $z = 249 \rightarrow 0$, with $1024^3$ gas elements and $1024^3$ dark matter particles. We also vary the sub-grid AGN feedback prescription using $50\ \hmpc$ comoving volumes, with $512^3$ gas elements and $512^3$ dark matter particles. In both cases, the mass resolution is $1.82 \times 10^7 \msolar$ for gas elements and $9.6 \times 10^7 \msolar$ for dark matter particles. The minimum adaptive softening length is $\epsilon_{{\rm min}} = 0.5\ \hkpc$, which is 0.5\% of the mean interparticle spacing between dark matter particles. We test resolution convergence using $25\ \hmpc$ comoving volumes: the first with $256^3$ gas elements and dark matter particles (the same mass resolution as the larger volumes), and a higher resolution run with $512^3$ gas elements and dark matter particles. The higher resolution run has 8 times the mass resolution of the fiducial volume ($2.3 \times 10^6 \msolar$ and $1.2 \times 10^7 \msolar$ respectively). Cosmological initial conditions are generated using \music \citep{hahn_2011} assuming a cosmology consistent with \cite{planck_collab_2016}: $\Omega_{M} = 0.3,\ \Omega_{\Lambda} = 0.7,\ \Omega_b = 0.048,\ H_0 = 68\ {\rm km\ s}^{-1}{\rm Mpc}^{-1},\ \sigma_8 = 0.82$\ and $n_s = 0.97$. Table \ref{tab:runs} summarises the simulations used in this work. 

\begin{table*}
\begin{center}
\begin{tabular}{lcccccc}
\hline
Simulation & Box size ($\hmpc$) & Nr. of particles & 
Stellar Feedback & AGN winds & Jets & X-ray heating \\
\hline
Simba $100\ \hmpc$ & 100  & $2\times 1024^3$ & \checkmark  & \checkmark  & \checkmark & \checkmark  \\
Simba $50\ \hmpc$ & 50  & $2\times 512^3$ & \checkmark  & \checkmark  & \checkmark & \checkmark  \\
Simba $25\ \hmpc$ & 25  & $2\times 256^3$ & \checkmark  & \checkmark  & \checkmark & \checkmark  \\
Simba $25\ \hmpc$ (Higher resolution) & 25  & $2\times 512^3$ & \checkmark  & \checkmark  & \checkmark & \checkmark  \\
No-Xray & 50  & $2\times 512^3$ & \checkmark  & \checkmark  & \checkmark  & \\
No-jet & 50  & $2\times 512^3$ & \checkmark  & \checkmark & & \\
No-AGN & 50  & $2\times 512^3$ & \checkmark & & & \\
No-feedback & 50  & $2\times 512^3$ & & & & \\
\hline
\end{tabular}
\caption{\simba\ runs used in this work.}
\label{tab:runs}
\end{center}
\end{table*}

% Cooling and heating
Radiative cooling and photoionisation heating are implemented using the \grackle-3.1 library \citep{smith_2017} in non-equilibrium mode, which offers two main advantages over the \grackle-2.1 library used in \mufasa that improve the accuracy of the baryonic thermal evolution. First, adiabatic and radiative terms are evolved together during the cooling time-sub-step, resulting in a more accurate and stable thermal evolution. Secondly, self-shielding is included self-consistently during the simulation run following the \cite{rahmati_2013a} prescription, in which the metagalactic ionising flux is attenuated depending on gas density and assuming a \cite{haardt_2012} spatially uniform ionising background. The neutral hydrogen content of gas particles is computed on the fly using the same self-shielding prescription.

% Star formation
As in \mufasa, star formation is modelled using an H$_2$-based \cite{schmidt_1959} law, using the sub-grid \cite{krumholz_2011} prescription to compute the H$_2$ fraction based on metallicity and local column density.
% ISM pressurisation
We apply artificial ISM pressurisation at a minimum level required to resolve star-forming gas, \Romeel{ applied above a hydrogen number density $n_H$ of $n_{th}>0.13~{\rm cm}^{-3}$, such that the temperature of this gas has a floor of}
\begin{equation}
    {\rm log} (T/{\rm K}) = 4 + \frac{1}{3}\log_{10} \frac{n_H}{n_{th}}.
\end{equation}
\Romeel{We define ISM gas eligible for star formation as having a temperature less than 0.5 dex above this temperature floor, along with $n_H>n_{th}$. Note that not all ISM gas is actively star-forming, as it must also have $H_2$; at low metallicity, the threshold to form $H_2$ can be at densities well above $n_{th}$.}

% Chemical enrichment
The chemical enrichment model tracks 11 elements (H, He, C, N, O, Ne, Mg, Si, S, Ca, Fe) from Type II supernovae (SNII), Type Ia supernovae (SNIa), and Asymptotic Giant Branch (AGB) stars, as described in \citet{oppenheimer_2008}. For SNII we use the \cite{nomoto_2006} yields, interpolated to the metallicity of the gas being enriched. For SNIa we use the \cite{iwamoto_1999} yields, assuming that each supernova yields 1.4$\msolar$ of metals. For AGB stars, we follow \cite{oppenheimer_2008}, using lookup tables as a function of age and metallicity, assuming a helium fraction of 36\%, a nitrogen yield of 0.00118, and a mass-loss rate computed from the \cite{chabrier_2003} initial mass function (IMF). We assume solar abundances from \cite{asplund_2009} to convert metallicities into solar units. 

% Dust
Metal enrichment from massive stars is partly locked into dust, modelled as a fraction of each gas element's metal mass that is passively advected along with the gas, as described in ~\citet{li_2019}. The dust prescription broadly follows that of \cite{mckinnon_2016}, with some improvements. Dust particles are grown by condensation and by accretion of metals via two-body collisions with gas elements, and are destroyed (mass and metals returned to gaseous phase) by collisions with thermally excited gas elements and via supernova shocks.

% Black hole growth and feedback
\simba's main improvement relative to \mufasa is its treatment of black hole growth and feedback. Black holes are seeded with $M_{\rm seed} = 10^4 \msolar$ into galaxies with a minimum stellar mass of $\gtrsim 10^{9.5} \msolar$ and grown self-consistently during the simulation. Growth of black holes is implemented via torque-limited accretion using the prescription of \cite{angles-alcazar_2017a} (based on \citealt{hopkins_2011a}) for cold gas with T $< 10^5$\ K, and via Bondi accretion \citep{bondi_1952} for hot gas with T $> 10^5$\ K. 

\simba's AGN feedback model has three mechanisms: 1) a radiative mode at high Eddington ratios ($\fedd$), 2) jet mode at low $\fedd$ and 3) X-ray feedback. Radiative and jet mode feedback are implemented kinetically in a direction $\pm$ the angular momentum of the inner disk. Radiative mode winds are ejected at the temperature of the ISM with a velocity that scales with the black hole mass \citep[approximately following][]{perna_2017}.  Jets begin when $\fedd < 0.2$, with a velocity boost that increases as $\fedd$ decreases, up to a maximum of 7000 km s$^{-1}$ at $\fedd = 0.02$. \Romeel{The kinetic AGN feedback velocity is thus}
\begin{equation}
    v_w = 500 + 500(\log \mbh-6)/3 + 7000 \log(\fedd/0.2)\;\;\kms.
\end{equation}
\Romeel{These outflows are decoupled for $10^{-4}t_H$, so they explicitly do not impact the ISM; at full jet velocity, they typically go several tens of kpc before recoupling.} 
The outflow mass loading factor is allowed to vary such that the momentum output of the black hole is $20L/c$, where $L$ is the bolometric AGN luminosity. Feedback due to X-rays from the accretion disk follows \cite{choi_2012} and only occurs in galaxies with $f_{\rm gas} < 0.2$ and for black holes with full-velocity jets.  These black hole accretion and feedback subgrid models reproduce very well the observed quenched galaxy fractions and galaxy--black hole correlations~\citep{dave_2019,thomas_2019} \Romeel{and the radio galaxy population~\citep{thomas_2021}.}

Star formation-driven winds are an important mechanism for enriching the CGM, so we describe them and their chemical enrichment more fully.  Massive stars drive galactic outflows through a combination of SNII, radiation pressure and stellar winds; we use a subgrid model to represent the net effect of these phenomena via two-phase kinetic feedback. Wind particles with velocity ${\bf v}$ and acceleration ${\bf a}$ are ejected in the direction $\pm{\bf v}\times {\bf a}$ (within the smoothing length), characterised by two free parameters: the mass loading factor $\eta$, and the wind speed $v_w$.  The scaling of mass loading factor with stellar mass is based on mass outflow rates from star forming regions in the Feedback In Realistic Environments \citep[FIRE,][]{hopkins_2014} simulations, computed by tracking individual particles in \cite{angles-alcazar_2017b}. The scaling is described by a broken power law at $M_0 = 5.2 \times 10^9\msolar$:
\begin{equation}\label{eq:eta_mstar}
    \eta(\mstar) \propto 
    \begin{cases}
        9\Big(\frac{\mstar}{M_0}\Big)^{-0.317},& \text{if } \mstar<M_0\\
        9\Big(\frac{\mstar}{M_0}\Big)^{-0.761},& \text{if } \mstar>M_0\\
    \end{cases}
    ,
\end{equation}
with a flat $\eta(\mstar)$ for objects with fewer than 16 star particles, to allow growth in poorly resolved galaxies that would otherwise struggle to grow due to excessive feedback. \Romeel{$\mstar$ is obtained via an on-the-fly friends-of-friends finder applied to ISM gas and stars.}

The wind velocity scaling with galaxy properties is also based on the FIRE simulations \citep{muratov_2015}, with an extra velocity kick $\Delta v(0.25R_{\rm vir})$ to account for the difference in radius between the wind launch site (typically well within the ISM) and the larger radius of $0.25R_{\rm vir}$ used in the FIRE scalings and further increased to account for hydrodynamic slowing:
\begin{equation}
    v_w = 1.6 \Big(\frac{v_{\rm circ}}{200 \kms}\Big)^{0.12} v_{\rm circ} + \Delta v(0.25R_{\rm vir}).
\end{equation}
The circular velocity $v_{\rm circ}$ is obtained using a baryonic Tully-Fisher-based scaling relation \Romeel{from $\mstar$. $\Delta v(0.25R_{\rm vir})$ is computed based on the gravitational potential difference of a wind particle's launch location versus $0.25R_{\rm vir}$, as described in \citet{dave_2016}.}  

To model the observed two-phase nature of galactic winds, 30\% of wind particles are ejected hot, with a temperature set by the supernova energy ($u_{\rm SN} = 5.165\times 10^{15}\ {\rm erg\ g}^{-1}$) minus kinetic energy, while the remaining particles are ejected at $T\approx 10^3$K. Ejected wind particles are hydrodynamically decoupled to avoid numerical inaccuracies arising from gas elements with high Mach numbers relative to their surroundings, and cooling is shut off to allow hot winds to deposit their thermal energy into the CGM. Outflowing wind particles are recoupled when 1) the particle's velocity is similar to its surroundings and its density is less than that of the ISM, or 2) a density limit of less than 1\% of the ISM threshold density is reached, or 3) a time limit of 2\% of the Hubble time at launch has passed.

Metal loading by SNII is modelled by allowing wind particles to extract some metals from their surroundings at launch time, enabling winds to transport additional metals into the CGM. The metallicity added to the wind particles is given by:
\begin{equation}
    dZ = f_{\rm SNII}\ y_{\rm SNII}(Z) / {\rm MAX}(\eta,1),
\end{equation}
where $f_{\rm SNII} = 0.18$ is the mass loss fraction due to SNII, $y_{\rm SNII}(Z)$ is the metallicity-dependent yield for each species, and $\eta$ is the mass loading factor. The metal mass is subtracted from the surrounding gas in a kernel-weighted manner; at all times metal mass is conserved.  Dust in wind particles is assumed to be destroyed if the particle is launched hot, but remains fully intact if launched cool; this carries some dust into the CGM.

%Long lived stars also contribute time-delayed stellar feedback in the form of SNIa and AGB stars. Following \cite{scannapieco_2005}, SNIa are modelled as a prompt component (which acts at the time of star formation and is treated in the same way as the kinetic SNII feedback) and a time-delayed component that begins after 0.7 Gyr. The rates for the two components are taken from \cite{sullivan_2006} and each SNIa adds a total energy injection of $10^{51}$ ergs to the surrounding gas. AGB stars are modelled assuming thermal equilibrium following \cite{conroy_2015}, with a wind velocity of 100 km s$^{-1}$. The AGB energy deposition rate is computed from the mass loss rate (derived assuming a \citealt{chabrier_2003} IMF and a \citealt{bruzual_2003} stellar population model) and the velocity difference between the star and the neighbouring gas elements.
% Energy and metal enrichment from SNIa and AGB stars act on the nearest 16 gas elements surrounding the given star particle in a kernel-weighted manner.

% Galaxy finder
Halos are identified on-the-fly during the simulation run using {\sc Subfind} as implemented within \gizmo.  We do not identify sub-halos, but instead identify galaxies in post processing. Galaxies are identified using a 6D friends-of-friends galaxy finder, using a spatial linking length of 0.0056 times the mean interparticle spacing (twice the minimum softening) and a velocity linking length set by the local dispersion. Galaxy finding is applied to all stars, black holes and gas elements with $n_H>n_{th}$.
%; additionally all gas elements with neutral hydrogen fraction $> 0.001$ are assigned to the galaxy to which they are most gravitiationally bound. 
Haloes and galaxies are cross-matched in post-processing using the \texttt{yt}-based \citep{turk_2011} python package \caesar\footnote{\url{https://caesar.readthedocs.io}}, which also computes many key properties including the stellar mass and star formation rates (SFR). The SFR is taken to the be sum of the instantaneous SFRs of all the gas particles in the galaxy.

\Romeel{There are some free parameters in \simba\ such as the AGN outflow wind speed and the SF outflow velocity prefactor (1.6).  These were adjusted primarily to reproduce the evolution of the stellar mass function at $z=2$ and $z=0$.  Additionally, black hole accretion has a free parameter associated with the efficiency of accretion onto the black hole, which was tuned to a value of 10\% to reproduce the amplitude of the $\mstar-\mbh$ relation.}

\section{CGM Mass and Metal Budget}\label{sec:mass_metal_budget}

\subsection{Classifying CGM Gas}

To begin, we examine the mass and metal budget of CGM gas in various phases within \simba\ around different galaxy types. Such mass and metal budgets are key physical properties that observations aim to quantify using CGM absorption~\citep{peeples_2014}. This will set the stage for understanding the absorption line statistics that probe these different CGM phases, and how these are impacted by various modeled physical processes.   We focus on $z=0$, but these CGM budgets do not vary significantly over the range of redshifts explored by COS-Halos and COS-Dwarfs.

We classify host galaxies and CGM gas as follows. We consider only central galaxies, as the CGM of satellite galaxies is not easily distinguishable from the CGM of their central galaxies. Instead of treating satellite galaxies separately, we simply include contributions from satellite galaxies in the mass and metal budget of central galaxies, as would likely be done observationally.  We consider the extent of the CGM to be within the galaxy's FOF halo \citep[but see e.g.][for alternative definitions]{prochaska_2011, wilde_2020}. \Romeel{We have verified that defining the CGM to be within R$_{200}$ instead results in very minor differences to the results below.}

For each central galaxy, we compute the baryon mass and metal mass of each component within its CGM: stars, ISM gas, dust, wind particles, and cool, warm, and hot CGM gas.  We define the wind phase as particles that are currently hydrodynamically decoupled from the surrounding gas (described in \S\ref{sec:sims}). As mentioned in \S\ref{sec:sims} ISM gas has hydrogen number density $n>n_{th}$; this includes all star-forming gas.   All remaining gas elements within the halo are considered part of the CGM, which is split into 3 temperature ranges: 'cool' gas dominated by photoionisation $T < T_{\rm photo}$, set at $10^{4.5}$~K; 'warm' gas above the photoionisation threshold but below half the virial temperature of the halo ($T_{\rm photo} < T < 0.5T_{\rm vir}$); and 'hot' gas above half the virial temperature, $T > 0.5T_{\rm vir}$. $T_{\rm vir}$ is calculated according to equation 4 of \cite{mo_2002}:
\begin{equation}\label{eq:tvir}
    T_{\rm vir} = 3.6\times 10^5 \bigg[\frac{V_c}{100\ {\rm km\ s}^{-1}}\bigg]^2{\rm K},
\end{equation}
where $V_c$ is the circular velocity of the halo. We separate the hot and warm gas in terms of the virial temperature in order to draw a distinction between thermally stable virialised gas, and thermally unstable warm gas in a transition phase between thermalisation and the photoionised regime.  We note that previous studies have typically taken this separation at a fixed temperature such as $10^{4.5}$~K or $10^{5}$~K~\citep[e.g.][]{ford_2013,peeples_2014}, but given that we will be examining trends over a fairly wide range of halo masses, this partitioning better captures the physical demarcations between thermally stable and unstable gas over a range of halo masses.  \Romeel{Equation~\ref{eq:tvir} drops below $T_{\rm photo}$ for $M_{\rm halo}\la 10^{10.6}M_\odot$, but in \simba\ these halos are below our resolution limit so do not factor into our analysis.}

\subsection{Mass Budget}\label{sec:mass_budget}

\begin{figure*}
	\includegraphics[width=0.95\textwidth]{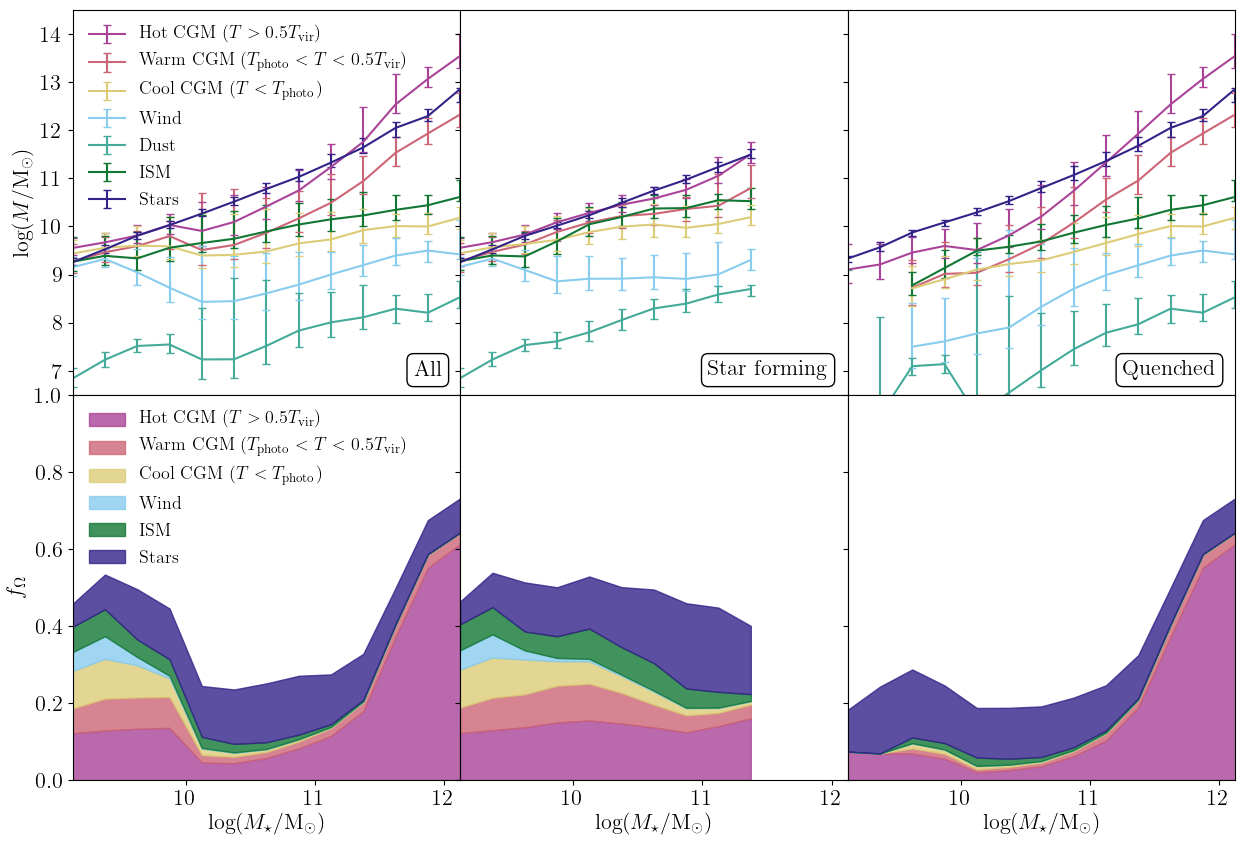}
	\vskip-0.1in
    \caption{Top panels: Median mass budget of different halo components at $z = 0$ (colour coded as in the legend inside the top-left panel), as a function of the stellar mass of the respective central galaxy. For every halo component, the error bars span the 25$^{\rm th}$-75$^{\rm th}$ percentile intervals of the mass distribution within each stellar mass bin. The left panel includes all galaxies, whereas the central and right panels are restricted to star forming and quenched galaxies, respectively. Bottom panels: Median mass fraction in each halo component (colour coded as in the upper panels), relative to the expected mass of cosmological baryons in each halo, as a function of central galaxy stellar mass. The three panels refer to all, star forming and quenched galaxies, as above. The baryon fraction is below unity at all masses, particularly for low mass quenched galaxies; quenched galaxy halos contain mostly hot gas while star forming galaxy halos are more diverse in gas phase.
    }
    \label{fig:mass_budget_omega}
\end{figure*}

Using these galaxy and phase classifications, we now examine the CGM mass budgets around central galaxies in \simba.  The upper panels in Figure \ref{fig:mass_budget_omega} show the median total mass of each halo baryonic component (hot CGM in magenta; warm CGM in peach; cool CGM in yellow; winds in light blue; dust in teal; ISM in green; stars in dark blue) for the halos of $z=0$ \simba\ central galaxies, as a function of central galaxy stellar mass. The lower panels show the median mass as a fraction of the baryon mass expected if halo has retained its cosmic share of baryons, defined as:
\begin{equation}
    f_\Omega = \frac{M}{M_{\rm halo}}\frac{\Omega_b}{\Omega_M}.
\end{equation}
The middle and right panels show the sample separated into star forming and quenched galaxies using a specific star formation rate (sSFR) cut of ${\rm log(sSFR / Gyr}^{-1}) > -1.8 + 0.3z$ as in \cite{dave_2019}. The error bars in the upper panels span from the 25th to 75th percentiles of the data.

In general each component mass increases as a function of stellar mass.  However relative contributions to the total mass budget vary substantially depending on the dominant physical mechanisms at play. At the high mass end, the trends are driven by galaxies that are quenched, primarily due to AGN jet feedback from massive black holes~\citep{dave_2019}, whereas at the low mass end galaxies are predominantly star forming since their AGN feedback is comparatively weak.  We examine this in more detail in \S\ref{sec:feedback}.

Broadly, across all stellar masses the largest contributions to the mass budget come from stars, hot CGM, warm CGM, ISM gas, and cool CGM,  while winds and dust contribute small amounts (see upper left panel).  At low masses, the various CGM phases have comparable mass, growing with $\mstar$ more slowly than the stellar mass.  At $\mstar\ga 10^{10.5}M_\odot$, the hot and warm CGM start increasing more quickly, while the cool CGM and ISM continue a nearly flat trend with $\mstar$.  At $\mstar\ga 10^{11}M_\odot$, the hot CGM mass begins to dominate over the stellar mass, leading to significant X-ray emission~\citep[e.g.][]{robson_2020}.  The low wind phase mass reassures us that measurements of metal-line absorption in our simulations are not significantly impacted by including hydrodynamically decoupled particles, except perhaps in the very lowest mass galaxies that have large mass loading factors. The dust mass contributes negligibly to the overall CGM mass budget, but may have interesting consequences for background source reddening~\citep{peek_2015}; we will explore this in future work.

These trends can be better understood by separating the galaxy population into star-forming and quenched galaxies (see the middle and right panels of Figure \ref{fig:mass_budget_omega}).  For star-forming galaxies, the hot CGM increases commensurately with the stellar mass, and is always comparable to it.  At low masses this likely owes to hot winds carrying hot material into the CGM, while at higher masses this owes to the onset of a shock-heated gaseous halo~\citep{birnboim_dekel_2003,keres_2005}, leading to a coincidentally steady trend with $\mstar$.  The ISM increases more slowly with $\mstar$, leading to lower galaxy gas fractions at higher masses~\citep{dave_2020}. The cool and warm CGM masses roughly track the ISM mass and are broadly comparable to it.  We note that the scaling of hot CGM with stellar mass owes to our choice of a halo mass-dependent division between hot, warm, and cool CGM gas; previous results using a constant temperature threshold tended to find that the hot component becomes small at low masses~\citep{keres_2009,faucher-giguere_2011,gabor_2015}, but this owes to the fact that the virial temperature becomes comparable or below what is canonically considered hot gas.

The quenched galaxies (rightmost panels), on the other hand, show significantly different trends particularly for the warmer CGM phases.  The hot and warm CGM masses increase dramatically with $\mstar$, from being nonexistent at $\mstar\la 10^{9.5}M_\odot$ to dominating the overall halo baryonic mass at $\mstar\ga 10^{11}M_\odot$.  This \Romeel{primarily owes to virial shock heating, exacerbated by} the onset of halo quenching owing primarily to jet AGN feedback in \simba, which enacts quenching by heating halo gas~\citep[e.g.][]{gabor_2015}.  The warm gas increases in tandem because as hot gas accumulates more of it is able to cool through the thermally unstable warm regime. Interestingly, the quenching does not fully remove the gas from the ISM and cool CGM phases; these continue to grow in mass even in hot halos at a similar rate as seen in star-forming galaxies.  It may be surprising to see that the halos of massive quenched galaxies contain a significant mass of cool CGM gas, hinting that the CGM of these quenched galaxies is multi-phase. To a lesser extent this is also true of ISM gas, although this gas is likely to be associated with satellite galaxies' ISM.

The wind mass for star-forming galaxies is roughly constant with stellar mass, reflecting a transition in the dominant wind outflow mechanism. The majority of the wind mass is linked to star-formation driven feedback, which has high mass and metal loading factors and long decoupling times to enable metal deposition in the CGM. Star-formation driven winds are most effective in low mass galaxies, resulting in high wind masses. Conversely, there is little mass in AGN-driven winds due to their relatively lower mass and metal loading factors and short decoupling times, leading to lower wind masses around quenched systems.

The bottom panels of Figure \ref{fig:mass_budget_omega} show that the overall baryon fraction is well below unity at all stellar masses, indicating that the halos have less mass in baryons than their cosmic share. This is a non-trivial result; in the simplest scenario, one expects that baryons would cool and condense into the central region of the halo, and the reduced pressure support would result in this fraction exceeding unity, which indeed happens in simulations without any feedback~\citep{dave_2009}.  The introduction of feedback results in significant fractions of the baryons being lost to the IGM. As noted in \cite{borrow_2020} and \citet{robson_2020}, \simba halos around $\sim L^\star$ galaxies struggle to retain more than 30\% of their baryons.

Star forming galaxies \Romeel{(lower middle panel)} are able to retain a higher fraction of their baryons, but even without the strong AGN feedback that would enact quenching, their halos still only retain around 40-50\% of their baryons. Examining a \simba run without jet-mode feedback shows that the halos of star forming galaxies retain a much higher fraction of their baryons, between $\sim 60\%$ for $\mstar \sim 10^{9.5}\msolar$ and $\gtrsim 90\%$ for $\mstar > 10^{11}\msolar$ (see \S\ref{sec:feedback}).  \Romeel{The stellar fraction continues to grow to the highest masses, consistent with observations of star-forming galaxies above $L^\star$~\citep{posti_2019}. \citet{cui_2021} has shown that in \simba, red and blue galaxies follow different stellar-to-halo mass relations that are individually in good agreement with data.}
This indicates that AGN jet mode feedback has two effects -- both evacuating baryons from their own host halos, and also reducing the overall fraction of baryons contained in any halo.
\mycolor{The overall baryon fraction in \simba follows a similar trend to IllustrisTNG, with a clear minimum around $L^\star$ systems, while EAGLE's baryon fractions stay more depressed to the lowest masses~\citep{davies_2019}.}

Quenched galaxies \Romeel{(lower right panel)} with $\mstar\la 10^{11}\msolar$ only contain $\sim 20\%$ of their share of baryons, indicating that feedback is highly effective at removing baryons and thus preventing accretion.  Feedback is most efficient in this mass range since the halos lack the gravitational potential to retain their baryons against the energetic input from AGN feedback. The jets could be quenching star-forming galaxies by heating and `boiling off' their gas supply, or by limiting accretion of new gas via preventive feedback \citep[e.g.][]{lu_2015, lu_2017, van_de_voort_2017}. The roughly constant fraction of baryons with stellar mass hints that it is the latter; if halos were losing gas due to heating we would expect an increase in baryon fraction in halos with higher gravitational potential. Galaxy clustering may compound this effect on higher mass galaxies (explaining the downturn at high mass), since these galaxies are likely to exist in high density regions \citep{gabor_2015,cucciati_2017, kraljic_2020} and hence are more likely to have neighbours with strong jet-mode AGN feedback.  \Romeel{We note that \citet{robson_2020} found good agreement with observed hot gas fractions as a function of halo mass at galaxy group scales in \simba, and demonstrated that this owes to jet feedback in \simba.}

\Romeel{Low-mass ($\mstar\la 10^{10}M_\odot$) quenched centrals} \response{could be influenced by environmental processes associated with large-scale structure or AGN feedback from nearby massive galaxies}; \simba reproduces the numbers of such \response{low-mass central} galaxies quite well~\citep[see figs. 4 and 5 in][]{dickey_2020}. Finally, we note the strong drop in the overall halo baryon fractions at $\mstar\sim 10^{10}M_\odot$, which occurs because the black holes here typically grow large enough to turn on jet feedback, resulting in the appearance of quenched galaxies~\citep{dave_2019}.  The drop is not evident in the star forming and quenched galaxy samples, but happens in the combined sample because the fraction of quenched galaxies jumps (perhaps too) abruptly at this mass.

In summary, the interplay of various feedback mechanisms results in a complex dependence of the halo baryon fraction on stellar mass in \simba, being most suppressed between $10^{10}\la \mstar\la 10^{11}M_\odot$ and higher outside these masses.  This is particularly evident when looking at quenched galaxies, which have a total halo baryon fraction of $\sim 20\%$ for $\mstar\la 10^{11}M_\odot$, indicating widespread evacuation of halos by feedback that enacts rapid quenching~\citep{rodriguez_montero_2019}. \citet{oppenheimer_2020} noted that CGM evacuation is also the cause of quenching in the EAGLE simulation.  Moreover, quenched galaxies' CGM are dominated by hot gas, although cool CGM gas has not completely vanished. Star-forming galaxies meanwhile have a multi-phase CGM, and show a fairly constant baryon fraction of $\sim 40-50\%$ at all masses.  Interestingly, there is always a substantial hot CGM component of $\sim 30-40\%$ of the total halo baryons, which indicates that virial-temperature gas is omnipresent even at low $\mstar$; the CGM of such galaxies at $z=0$ is not completely dominated by cold infalling filaments as has been argued from previous simulations~\citep{keres_2005,dekel_2009}.

\subsection{Metal Budget}

\begin{figure*}
	\includegraphics[width=0.95\textwidth]{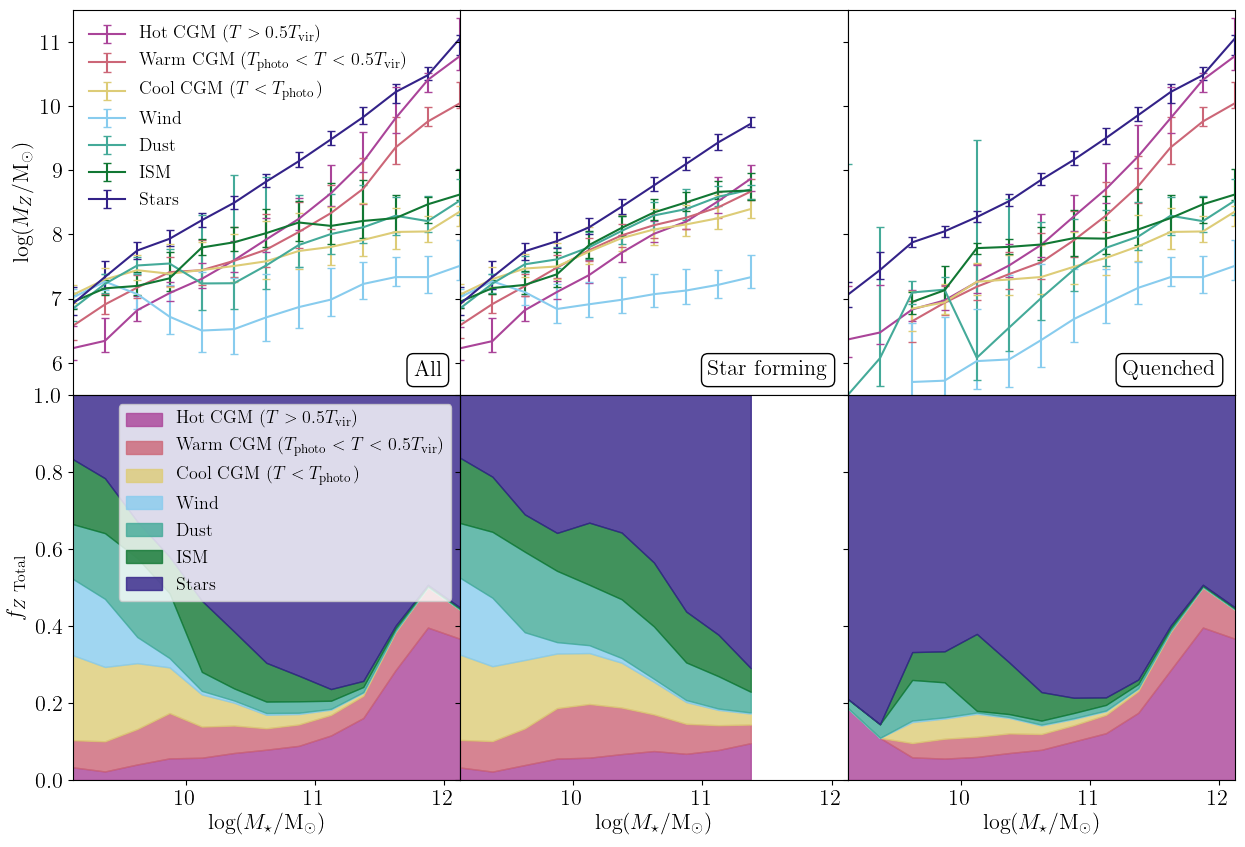}
	\vskip-0.1in
    \caption{Top panels: As in Figure~\ref{fig:mass_budget_omega}, except that we now show the median metal mass of each halo component. Bottom panels: Median metal mass fraction of each component relative to total available metal mass, following the same colour coding and panel structure as in Figure~\ref{fig:mass_budget_omega}. Quenched galaxies have most of their metals locked into stars, whereas in star forming galaxies no single component dominates the metal mass fraction.
    }
    \label{fig:metal_budget}
\end{figure*}

Since CGM tracers are predominantly metal absorption lines, it is instructive to examine the metal content of halos within the various CGM phases and galaxy types. Figure \ref{fig:metal_budget} shows the median metal mass (upper) and metal mass fraction (lower) for each halo component (colours same as Figure \ref{fig:mass_budget_omega}), as a function of central galaxy stellar mass. The error bars in the upper panels indicate the 25th and 75th percentiles of the data. Galaxies are separated into star forming and quenched using the same sSFR cut as in Figure \ref{fig:mass_budget_omega}. 

At the low mass end, metals are evenly distributed across stars, ISM gas, wind particles, dust and cool CGM gas, with $\sim 10\%$ of metals in warm and hot CGM gas. Compared with their overall mass, warm and hot CGM gas contributes relatively little to the metal budget. At intermediate masses, metals are more preferentially locked into stars, and at the high mass end the metal mass of the hot CGM gas increases to $\sim 35\%$ of the available metals; both of these effects are due to the increasing mass of these components overall (see Fig. \ref{fig:mass_budget_omega}). 
Dust makes a significant contribution to the metal mass, comparable to the ISM as found in \citet{li_2019}.  \Romeel{The large majority of dust is in the ISM; the dust ejected in (cool) outflows is highly uncertain since the production and destruction mechanisms under such conditions are not well constrained.}
The metal mass fraction in winds is strongly dependent on stellar mass: in galaxies with $\mstar > 10^{10} \msolar$ winds compose a negligible proportion of the metals, whereas low mass galaxies have $\sim 20\%$ of their metal mass in wind particles, owing to the increase mass in winds and more substantive metal loading.

Star forming galaxies have a higher fractions overall of metals in dust, the ISM and cool CGM gas. The metal mass in stars increases steeply with stellar mass such that at the high mass end most metals are locked into stars, with the remaining metal budget composed equally of ISM gas, dust, and cool, warm and hot CGM gas. At low masses, star forming galaxies have a metal mass in dust that is comparable to that in stars, owing to the high mass overall of metal-loaded wind particles. Wind metal mass is roughly constant with stellar mass, such that the fraction at the high mass end is negligible.  Overall, no single halo gas phase dominates the metal budget in star-forming systems.

Quenched galaxies have most of their metals locked into stars at moderate stellar masses. The higher fraction of metals in stars is due to the lower metal mass in other components as compared to star-forming galaxies; the metal mass locked into stars is similar at a given $\mstar$. The significant proportion of metals in warm and hot CGM gas is in line with the overall gas masses of these components. Metals could end up in these hotter phases by direct transportation of heated gas from the galaxies via AGN jets, or via heating by e.g. virialization or hot jet outflows in the CGM following ejection of cooler star formation driven winds; an investigation into how metals arrive in this phase is left for future work utilising particle tracking.

\Romeel{
\citet{peeples_2014} conducted a halo metal census inferred from ionisation modeling of COS-Halos metal lines and other data.  Our results on the relative amounts of metals in various phases are in fairly good agreement with this.  Generally, \citet{peeples_2014} found that the fraction of metals in stars increases with $\mstar$, in ISM gas decreases with $\mstar$, and in dust is fairly constant; these qualitatively match \simba predictions.  The CGM metal fractions were less well constrained, but overall they found that it increased to lower masses, which \simba likewise predicts.  \simba results are thus broadly consistent with this more model-independent metal census.
}

\begin{figure*}
	\includegraphics[width=0.95\textwidth]{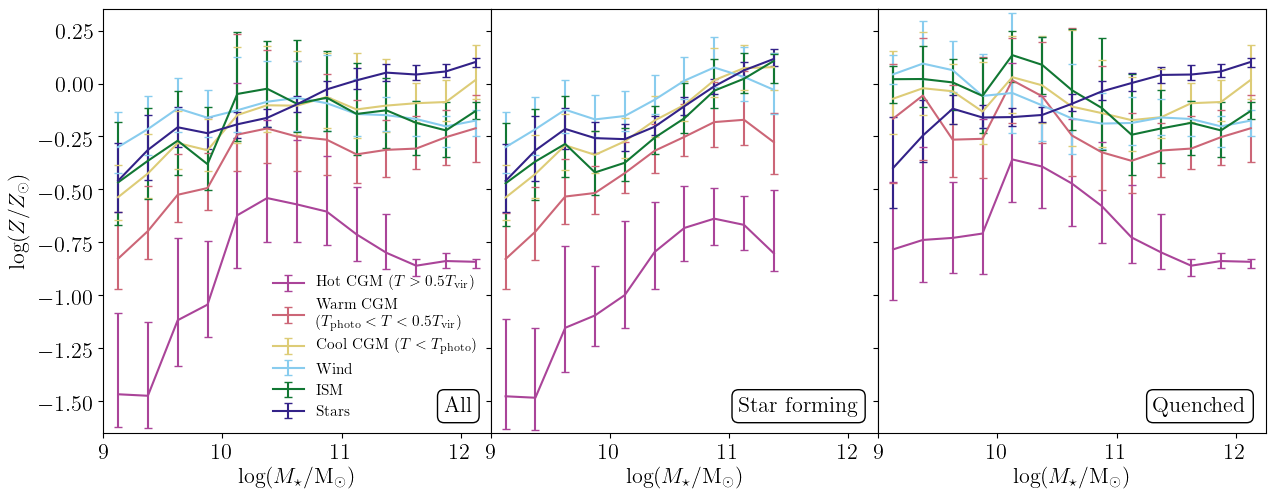}
	\vskip-0.1in
    \caption{Median metallicity of each halo component as a function of central galaxy stellar mass, separated into all (left), star forming(middle) and quenched (right) galaxies. The meaning of the error bars and the colour coding is the same as in Figure~\ref{fig:mass_budget_omega}. For quenched systems the CGM metallicity is roughly constant with $\mstar$, while for star forming systems it increases with $\mstar$.
    }
    \label{fig:metallcity}
\end{figure*}

While the metal mass is useful for budgeting, the metallicity is an indicator of the physical origins of the gas; low metallicity, `pristine' gas is likely to have been accreted from the IGM, whereas higher metallicity gas has been enriched by star formation in galaxies, hence the mass-metallicity relation places constraints on galactic outflows \citep{finlator_2008}. \Romeel{\citet{dave_2019} showed that \simba\ generally reproduces the observed galaxy gas-phase metallicity--stellar mass relation at both $z=0$ and $z\approx 2$; here we examine the metallicity of gas that has ended up outside the ISM.}

Figure \ref{fig:metallcity} shows the median metallicity of each halo component as a function of central galaxy stellar mass, separated into star forming and quenched galaxies. Metallicity is computed as the total metal mass fraction of each component $i$:
\begin{equation}
    Z_i = \frac{M_{Zi}}{M_{{\rm gas}\ i}},
\end{equation}
scaled by the solar metallicity $Z_{\odot} = 0.0134$ \citep{asplund_2009}. Dust metallicity is not included in this figure since by definition dust is composed 100\% of metals. 
The trends in stellar and ISM gas-phase metallicity with stellar mass represents the well-known mass-metallicity relation~\citep[e.g.][]{tremonti_2004}; see \citet{dave_2019} for a detailed discussion of these relations in \simba.  Here we go further by decomposing the total gas into distinct CGM components.

The ISM and CGM components tend to show a rise in metallicity at low masses ($\mstar\la 10^{10.5}M_\odot$), and then a flat or even downwards trend towards higher masses, in contrast to stellar metallicities which are continually increasing.  In \citet{finlator_2008}, the galaxy metallicity scales as the yield divided by $(1+\eta)$, where $\eta$ is the mass loading factor.  Thus for the ISM, this flattening trend can be understood as $\eta$ approaching unity at $\mstar\ga 10^{10.5}M_\odot$.  The CGM components, interestingly, reflect this ISM trend, which indicates that the CGM metals are enriched by the winds from the galaxy.  

The hot CGM shows the lowest metallicity among all the halo gas components, indicating the highest amount of less enriched infall diluting the CGM contributions from outflows.  The hot gas that is enriched is more likely to suffer metal line cooling and move to lower-temperature phases, which selects the lowest metallicity gas to remain hot.  The increased halo baryon mass fraction at the highest masses along with the dropping metallicity suggests that at high $\mstar$ the galaxies' halos are preferentially able to retain their cosmic inflow against feedback, which is as expected owing to their larger potential wells.

Examining the star forming and quenched populations separately reveals an interesting dichotomy.  For star-forming galaxies, the CGM uniformly increases in metallicity in all phases, except for the few very most massive systems, indicating continual enrichment up to the highest masses.  In contrast, the quenched galaxy CGM metallicities are remarkably constant, though they show a mild peak at $\sim 10^{10.3}M_\odot$.  These galaxies are not forming new metals, so the CGM of these systems must have been enriched during the star-forming phase, or by accreting metals from other galaxies~\citep{angles-alcazar_2017a}.  The hot phase metallicity is $\sim 1/4$ solar, which is broadly as observed in X-ray gas in groups and clusters, but shows a trend of increasing metallicity towards lower $\mstar$ (and thus lower $M_{\rm halo}$) systems that is an interesting prediction.  The other components tend to be around solar to half-solar.  

Overall, the metal mass budget of the CGM partly traces the gas mass budget, but is impacted by the different origin of gas around star-forming and quenched systems.  Stars generally dominate the overall budget of metals remaining in the halo, except at the lowest masses probed here.  The metals are generally multi-phase (like the mass) in star-forming systems, but in quenched systems the cooler components are important at low masses but the hot gas dominates the metal budget at $\mstar\ga 10^{11}M_\odot$.  
\Romeel{Many of these metals were likely accreted from the IGM at late epochs~\citep{oppenheimer_2012}, and some may have been deposited by satellites via stripping or outflows; we leave an investigation on the origin of these metals for future work.}
The metallicities of these components also show stark differences between star-forming and quenched systems, with star-forming galaxies having increasing metallicities with $\mstar$ in all components, while quenched systems have approximately constant metallicities at all $\mstar$, apart from in the stellar component.  The low metallicity ($\sim 10\%$ solar) of the hot CGM is a reflection of both less enriched infall and metal cooling.  These trends are predictions from \simba that should in principle be decipherable from a suite of well-characterised CGM absorbers.

\section{Comparison to COS-Halos and COS-Dwarfs}\label{sec:cos_comparison}

We now examine the CGM in absorption in \simba. We focus on recent absorption-line studies using the Cosmic Origins Spectrograph (COS) probing the CGM of \lstar galaxies \citep[COS-Halos,][]{werk_2013, werk_2014, tumlinson_2013} and sub-\lstar dwarf galaxies \citep[COS-Dwarfs,][]{bordoloi_2014}. In order to test how well the CGM might be represented in \simba, we perform a close comparison to these observations by examining the CGM in \simba in terms of observational absorption statistics. We conduct analogous absorption-line surveys within the \simba volume that mimic the selection properties and instrumental characteristics of these observational surveys.

\subsection{Mocking the COS surveys}

\subsubsection{Galaxy selection}\label{sec:galaxy_selection}

\begin{figure}
	\includegraphics[width=\columnwidth]{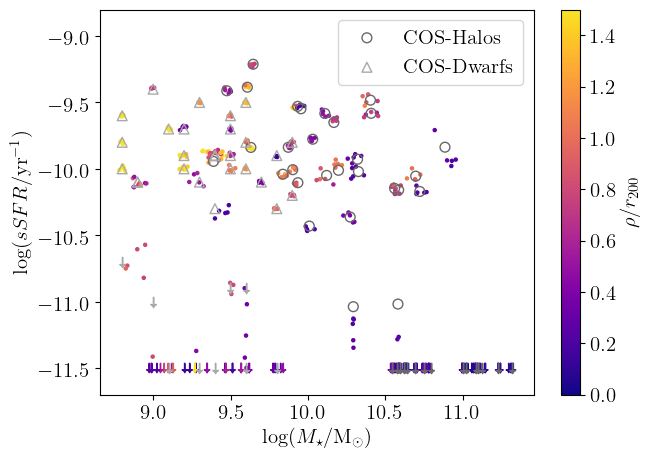}
	\vskip-0.1in
    \caption{Specific star formation rate (sSFR) versus stellar mass (\(\mstar\)) for the \simba galaxies in our COS-Dwarfs and COS-Halos samples. Points are colour-coded by their $r_{200}$-scaled impact parameter. Dark grey circles and light grey triangles represent galaxies in COS-Halos and COS-Dwarfs, respectively. Where sSFR is a lower-limit for these galaxies, this is indicated by a downward arrow. Where simulated galaxies have little or no star formation we set their sSFR to \(10^{-11.5} \textrm{yr}^{-1}\) for plotting purposes, also indicated by downward arrows.
    }
    \label{fig:galaxy_samples}
\end{figure}

To compare to these observations of the CGM in absorption, we choose galaxies from the fiducial \simba run whose properties are similar to those of the surveyed galaxies. \Romeel{Overall, our procedure for this closely follows that in \citet{ford_2016}. Other CGM absorption studies have compared specifically to COS-Halos,} \response{either by similarly creating matched galaxy samples \citep{oppenheimer_2016, nelson_2018b} or neglecting this extra step of observational realism \citep[e.g.][]{hummels_2013, ford_2016}}.

For each galaxy in the observational sample, we select the 5 galaxies in \simba with the closest match in terms of stellar mass and sSFR. \mycolor{We scale stellar masses in COS-Dwarfs and COS-Halos to the Chabrier IMF assumed in \simba\ by dividing by a factor of 1.6~\citep{madau_2014}.}  We impose an isolation criteria to mimic that of the COS surveys, requiring that no other central galaxies are within 1 $\hmpc$ of each selected galaxy. 

For the COS-Dwarfs and COS-Halos samples we select galaxies from the $z=0$  and $z=0.25$ snapshots, respectively, closest to the typical redshifts of these surveys. We consider only galaxies with $\mstar > 5.8 \times 10^8 \msolar$, which is \simba's nominal mass resolution limit, meaning we exclude the lowest mass galaxies in COS-Dwarfs. As in \S\ref{sec:mass_metal_budget}, we then separate star forming and quenched populations using a specific star formation rate cut of ${\rm log(sSFR / Gyr)}^{-1} > -1.8 + 0.3z$. For COS-Dwarfs this gives \mycolor{29} star forming and 8 quenched galaxies in the original sample, and \mycolor{142} star forming and \mycolor{43} quenched analogous \lstar galaxies in \simba (where the non-exact split is due to a COS-Dwarfs galaxy that is near to our sSFR boundary).  For COS-Halos, this gives 27 star forming and 17 quenched galaxies in the original sample, and 135 star forming and 85 quenched analogous sub-\lstar galaxies in \simba.

Figure \ref{fig:galaxy_samples} shows stellar mass against sSFR, colour-coded by impact parameter $\rho$ scaled to $r_{200}$ of the host galaxy's halo, for galaxies in our mock COS-Halos and COS-Dwarfs samples. \simba galaxies with little or no star formation are plotted as downward arrows at an sSFR of \(10^{-11.5} \textrm{yr}^{-1}\); \Romeel{for matching purposes, these are treated at having an sSFR of \(10^{-11.5} \textrm{yr}^{-1}\)}. The original survey galaxies are also shown as dark grey circles (COS-Halos) and light grey triangles (COS-Dwarfs). Grey arrows indicate lower-limit sSFRs for some observed galaxies.  \Romeel{This shows that our simulated galaxies are a good representation of the COS samples.}

\subsubsection{Spectral generation}

For each galaxy in our sample, we select lines of sight (LOS) through the simulation to probe its CGM at the impact parameter $\rho$ of the corresponding COS galaxy-quasar pair. We choose 8 LOS per galaxy with impact parameter $\rho$, starting from ($x_{\rm gal}+\rho, y_{\rm gal}$) and proceeding anti-clockwise, selecting a point every 45 degrees. All LOS are parallel to the z-axis of the simulation box. We generate absorption spectra for a selection of ions which probe a range of ionisation potentials that are covered by the COS surveys: \HI 1215\AA\ and \CIV 1548\AA\ for the COS-Dwarfs sample; \HI 1215\AA, \MgII 2796\AA, \SiIII 1206\AA, and \OVI 1031\AA\ for the COS-Halos sample.  These are among the most commonly seen lines in these surveys, offering the largest statistical samples for comparison to \simba. When comparing \HI absorption we ignore one COS-Dwarfs galaxy with an intervening Lyman limit system. 

Mock absorption spectra are generated through the simulation volume using the spectrum generation tools in the \pygad\footnote{https://bitbucket.org/broett/pygad} analysis package \citep{rottgers_2020}.\mycolor{The spectra are computed from gas particles who smoothing lengths intersect with the LOS, so spectrum generation is not dependent on our halo-finding procedure.} The ionisation fractions for each species are found using look up tables in density and temperature generated using version 17.01 of \cloudy \citep{ferland_2017}. Self-shielding is applied for metal lines via an attenuation of the overall background following \cite{rahmati_2013a} prescription\footnote{This is applied as an overall reduction factor to the entire ionising background spectrum in the optically thin limit; this is not correct in detail since self-shielding also changes the spectral shape, but it is the best we can do in lieu of an expensive multi-frequency radiative transfer calculation.}.
\Romeel{The amount of neutral \ion{H}{i} is calculated directly during the simulation run, including self-shielding effects, so we do not re-derive the ionization state of hydrogen in post-processing.}
The particle densities are multiplied by the ion fraction for each species, assuming a $\approx$76\% mass fraction for H, and the individually-tracked mass fractions within \simba\ for each metal. The resulting ion densities are smoothed along the LOS using the same spline kernel used in the \gizmo\ simulation code, employing the smoothing length and (for metal ions) the relevant metal content of each particle, into 6$\kms$ width pixels (approximately corresponding to COS pixels) along the LOS. The resulting ion column densities for each pixel are converted into an optical depth using the oscillator strength of each species.  See \citet{rottgers_2020} for further details on \pygad.

We add effects to the spectra to approximately mimic the COS observations: i) spectra are convolved with the line spread function for the COS G130M grating (apart from the \MgII spectra, whose observations were taken with Keck; \citealt{werk_2013}); ii) random noise is added with a signal-to-noise ratio (SNR) per pixel of 12, corresponding to SNR per COS resolution element of $\approx 20$; iii) we mock a continuum fitting procedure broadly following \cite{danforth_2016} by iteratively removing pixels that are $> 2\sigma$ below a polynomial best-fit to the spectrum until the continuum fit changes fractionally by less than $10^{-4}$.

\subsubsection{Choice of photo-ionising background}

Computing the ionisation fractions in \cloudy requires assuming a spectrum of incident photons, which is critical since most of the CGM UV absorption lines detected are typically photo-ionised~\citep{ford_2013}, potentially even \OVI~\citep{oppenheimer_2009a}. We find that our choice of photo-ionising background can make a significant difference to the metal line absorption statistics; unfortunately this choice remains rather poorly constrained by direct observations. 

We will thus compare results for three different ionising backgrounds: (i)  \cite{faucher-giguere_2020} (hereafter \citetalias{faucher-giguere_2020}), an updated version of the \citep{faucher-giguere_2009} background; (ii) \citet{haardt_2001} (hereafter \citetalias{haardt_2001}), an older determination utilised in many previous studies including \citet{ford_2013}; and (iii) \cite{haardt_2012}, with the entire spectrum increased in amplitude by a factor of 2 (hereafter \citetalias{haardt_2012}).

The low-redshift \HI\ photoionisation rate can be constrained by comparing to the observed \HI\ column density distribution~\citep[e.g.][]{kollmeier_2014} or mean flux decrement~\citep[e.g.][]{christiansen_2020}.  For \simba, the mean flux decrement is reasonably well-matched to data assuming \citetalias{faucher-giguere_2020}, whereas the HM12 model requires an increase of a factor of 2.  This gives a $z=0$ \HI\ photoionisation rate of $\ghi=4.7\times 10^{-14}$ and $\ghi=4.5\times 10^{-14}$s$^{-1}$ for our two backgrounds, respectively, so differences in absorption properties will reflect variations in the shape of the background above 1~Ry.  \citetalias{haardt_2001} is higher in terms of the \HI photoionisation rate ($\ghi=8.3\times 10^{-14}$s$^{-1}$ at $z=0$), and its shape is substantially different owing to a larger assumed galaxy escape fraction and not including a soft X-ray background; we include it here for comparison to previous results.

\Romeel{We note that \simba\ was run with the original \citet{haardt_2012} background.  Therefore none of the choices here are self-consistent with the simulation run.  Fortunately, the ionising background makes a negligible difference in the hydrodynamical forces in cosmological simulations, so applying the correction in post-processing is a very good approximation~\citep{weinberg_1997}.}

\subsubsection{Absorption statistics}

From the synthetic spectra we compute 3 observables in order to compare with the COS surveys: (i) the equivalent width EW, computed within a $\pm 300 \kms$ window centered on each galaxy; (ii) the covering fraction \fcov, which is the fraction of lines of sight where the equivalent width exceeds the detection threshold; and (iii) the path absorption $dEW/dz$, which is the cumulative equivalent width in $\pm 300 \kms$ per unit redshift.

The EW is the most basic statistic for unresolved or marginally resolved absorbers, which is typically the case for metal lines from cooler gas observed with COS.  For \fcov, we employ a detection threshold of 0.2~\AA\ for \HI and 0.1~\AA\ for all metal ions, following \citet{werk_2013} and \citet{ford_2016}.  The covering fraction constrains the spatial structure and extent of the CGM absorption in each ion.  The path absorption is effectively a stacked measure of circumgalactic absorption, computed directly from the spectra without any consideration of individual lines or detection thresholds.  These measures thus provide differing quantifications of CGM absorption.

To compare to COS-Halos, we access the data from the \pyigm\footnote{\url{https://pyigm.readthedocs.io/en/latest/}} repository, which includes the equivalent widths, column densities and Doppler parameters for each metal line component of the absorption spectra. \Romeel{This repository compiles the data from the COS-Halos and COS-Dwarfs surveys with {\it Hubble}, along with ancillary data including \ion{Mg}{ii} from Keck spectra and galaxy properties from spectroscopic follow-up.  The data are reduced and analysed as described in \citet{tumlinson_2013,werk_2014}, and the analysis tools are available on the repository (though here we rely on the reduced data products).} For COS-Dwarfs, the \HI\ and \CIV\ data as a function of impact parameter was kindly provided to us by survey PI R. Bordoloi.

We focus on the EWs here, since in many cases the Doppler parameter cannot be directly determined from the spectra, and the \pyigm column densities therefore assume a Doppler parameter that may not be correct~\citep{rottgers_2020}.  \Romeel{The EW in \pyigm were measured in $\pm 600\kms$ windows around the galaxy redshift, after continuum fitting.  In our case, we measure EW in a smaller window, but have checked that the contribution from $\pm 300-600\kms$ is negligible, as also found in \citet{ford_2016}. We consider all gas along the line of sight through the entire volume that contribute within this velocity window; some of it may lie outside of halos and be projected within in redshift space~\citep{rahmati_2015,nelson_2018b}.}

\begin{table}
\caption{Ion Properties and detection thresholds}
\begin{tabular}{lccc}
\hline
Ion & Ionisation Energy (Ry) &
Detection Threshold (\AA)
\\
\hline
\multicolumn{3}{c}{}\\
\HI\ & 1.0 & 0.2 \\
\ion{Mg}{ii} & 1.11 & 0.1 \\
\ion{Si}{iii} & 2.46 & 0.1 \\
\ion{C}{iv} & 4.71 & 0.1 \\
\ion{O}{vi} & 10.15 & 0.1 \\
\hline
\end{tabular}
\label{table:thresh}
\end{table}

\Romeel{As in \citet{ford_2016}, we set a fixed EW detection threshold for each ion, motivated by the sensitivity limits of the COS surveys.  The detection limits are listed in Table~\ref{table:thresh}, and for reference the ionisation potentials are listed to show the range from low ions (\ion{Mg}{ii}) to high (\ion{O}{vi}).
We apply the same limits to the simulated absorbers.  Upper limits are treated as detections at that limit, but since we consider medians here, this only affects the results when there are very few detections.}

We choose to compare all spatial statistics with impact parameters normalised to $r_{200}$, defined as the radius enclosing 200 times the critical density.  We do this in order to be able to meaningfully compare profiles across the large mass range spanned by both COS-Halos and COS-Dwarfs, under the expectation that the influence of the baryon cycle around each galaxy is approximately proportional to the virial radius.  There is evidence for such self-similarity at least in terms of \MgII\ absorbers~\citep{churchill_2013}. In Appendix \ref{sec:kpc_units} we show the EW versus physical impact parameter instead, but this choice does not impact our results; it does however make some of the radial trends less evident, as also noted in \citet{ford_2016}.

For \simba\ we obtain $r_{200}$ directly from the identified halos for each (central) galaxy.  For the observations, $r_{200}$ was computed in \citet{werk_2013} from the stellar mass based on the $\mstar-M_{\rm halo}$ relation of \citet{moster_2010}.  We note that our simulations are in good agreement with the observed $\mstar-M_{\rm halo}$ relation for massive halos where this is observed via e.g. weak lensing, even when considering quenched and star-forming galaxies separately \citep{cui_2021}.

\subsection{Comparison to COS-Halos and COS-Dwarfs}

We now compare \simba\ to COS-Halos and COS-Dwarfs using the absorption spectra and statistics described above.  This section represents the main results of our paper.

\subsubsection{Equivalent widths}

\begin{figure*}
	\includegraphics[width=0.95\textwidth]{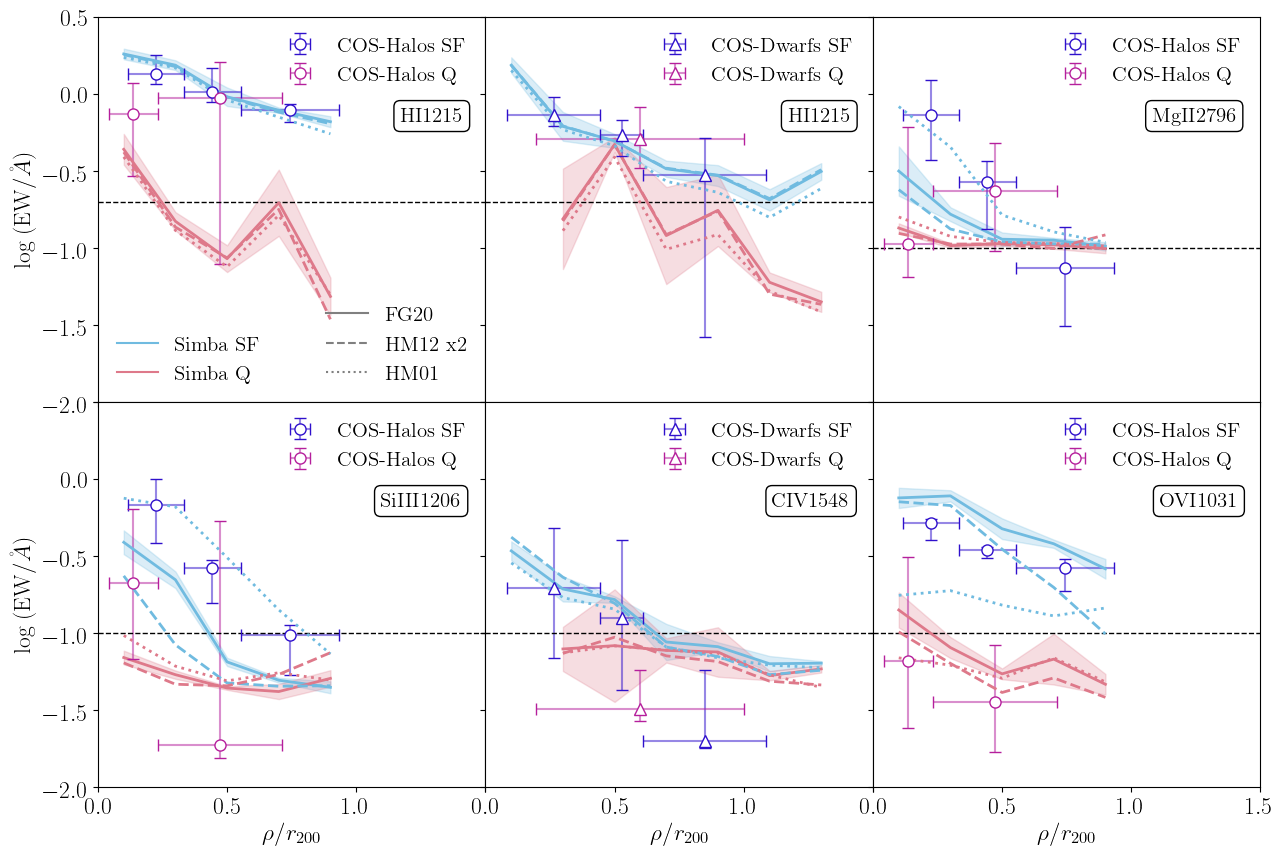}
	\vskip-0.1in
    \caption{Equivalent width of \HI and selected metal lines against (as indicated in each panel) $r_{200}$-scaled impact parameter for the COS-Halos and COS-Dwarfs galaxy samples, using the \protect\citetalias{faucher-giguere_2020} (solid), \protect\citetalias{haardt_2012} (dashed) and \protect\citetalias{haardt_2001} (dotted) ionising backgrounds. Light blue and light pink lines represent star forming and quenched galaxies in the \simba sample, respectively; shaded regions show indicative cosmic variance uncertainties around the \protect\citetalias{faucher-giguere_2020} results. Dark blue and magenta points represent star forming and quenched galaxies in the COS samples (circles for COS-Halos, triangles for COS-Dwarfs), respectively; vertical error bars represent the 25th and 75th percentiles of the equivalent width distribution in the data, while horizontal error bars indicate the width of the bins. Black horizontal dotted lines indicate the detection threshold of each line under the COS survey conditions. CGM absorption is higher around star forming galaxies than quenched galaxies. \simba is in reasonable agreement with observations; \HI absorption around star forming galaxies is well-reproduced, although metal line absorption is sensitive to the assumed ionising background.
    } 
    \label{fig:rho_ew}
\end{figure*}

Figure \ref{fig:rho_ew} shows median equivalent width of the selected ions, against $r_{200}$-scaled impact parameter for the COS-Halos and COS-Dwarfs galaxy samples, using the \citetalias{faucher-giguere_2020} (solid), \citetalias{haardt_2012} (dashed) and \citetalias{haardt_2001} (dotted) ionising backgrounds. \simba\ star forming and quenched galaxies are represented by light blue and light pink lines, respectively, and shaded regions indicate the cosmic variance within eight sub-quadrants over the simulation volume for the \citetalias{faucher-giguere_2020} results \Romeel{(which typically dominates over Poisson variance)}. Dark blue and magenta points represent the actual COS-Dwarfs/COS-Halos data, respectively; the error bars are the 25th and 75th percentiles of the data. The varying bin widths for the observational data are chosen such that each bin contains $\sim8$ galaxies.  The detection thresholds for each ion are indicated by horizontal dotted lines.  

% Hydrogen
In general, for each ion and for both star forming and quenched populations, the equivalent width of absorption is higher around star forming galaxies, and decreases with impact parameter. This latter feature is enhanced by scaling impact parameter by $r_{200}$, as can be seen by comparing to Figure \ref{fig:rho_ew_unscaled} in Appendix \ref{sec:kpc_units}, suggesting that CGM properties correlate better with $r_{200}$-scaled radii than physical radii.  \HI\ absorption is $\sim 10$ and $\sim 3-4$ times higher in \simba star forming galaxies than in quenched galaxies, for the mock COS-Halos and COS-Dwarfs samples respectively (upper left and upper middle panels). The predicted \HI\ absorption around star forming galaxies in \simba is in good agreement with both observational surveys. The accurate prediction of \HI\ absorption around star-forming galaxies is a non-trivial success that did not require any specific tuning. 

%{\bf Referring to Dave 2020 - At z=0 we agree well with HIMF compared with ALFALFA (slightly overproducing at the high mass end and for low sSFR), and overproduce for the COLF (molecular), but what about for intermediate and low mass quenched galaxies? The mass budgets of these systems (figures 2 and 3) show that at these masses the CGM has very little cold gas, which leads to the low absorption in quenched galaxies.}

In contrast, \simba's quenched galaxies show significantly less \HI\ absorption compared with the observations. Quenched galaxies in \simba contain little cold gas in their halos (see Figure \ref{fig:mass_budget_omega}) since the quenching mechanism heats and removes the cold gas supply and prevents further accretion via jet feedback. Meanwhile the COS-Halos observations suggest that real galaxies are quenched in such a way that they retain substantial cool gas in their CGM.  
The presence of cool gas around quenched galaxies remains an interesting and non-trivial quandary~\citep{thom_2012,tumlinson_2017} to constrain quenching models, though more data is needed to increase the statistical significance. 

Comparing COS-Halos \HI\ (upper left) with COS-Dwarfs \HI\ (upper middle), we see that there is substantially less separation between the star-forming and quenched absorption in the lower-mass sample.  This suggests that such lower mass galaxies are perhaps quenched by a different physical process.  In \simba, the AGN jet feedback that quenches galaxies only turns on at $\mstar\ga 10^{10}M_\odot$, so most COS-Dwarfs galaxies would not have this.  They may instead be ejected satellites or ``neighborhood quenched" galaxies~\citep{gabor_2015}, although the isolation criterion used in observations that we mimicked tried to avoid this.  We note that \citet{dickey_2020} examined the quenched fraction of dwarf galaxies in \simba\ compared to observations, and although there is an excess of quenched systems at the very lowest masses, in the COS-Dwarfs mass range \simba\ well reproduces the observed fraction of quenched dwarfs. 

Choice of photo-ionising background has little impact on our \HI\ absorption results. Assuming instead the \citetalias{haardt_2012} background results in almost identical \HI absorption around star forming galaxies in the COS-Halos and COS-Dwarfs samples, while the \citetalias{haardt_2001} background results in slightly lower \HI absorption. Around quenched galaxies in the COS-Halos and COS-Dwarfs samples, the \citetalias{haardt_2012} and \citetalias{haardt_2001} backgrounds both result in slightly higher \HI absorption than the \citetalias{faucher-giguere_2020} background. These are small differences within the uncertainties; \HI absorption is a less sensitive probe of ionisation variations due to the saturation of absorption features which places them on the logarithmic part of the curve of growth.

% Metals
Cool CGM gas can also be explored via low ionisation lines such as \MgII\ (upper right panel) and \SiIII\ (lower left panel).  In this case, the story is less conclusive. For the \citetalias{faucher-giguere_2020} and \citetalias{haardt_2012} backgrounds used here, \simba underpredicts the EWs of these low ionisation lines around star-forming galaxies by $\sim\times 3$. However, assuming \citetalias{haardt_2001}, \simba\ shows much better agreement. This highlights the sensitivity of the assumed ionising background shape in comparing to metal lines. Furthermore, there are uncertainties in the Mg and Si yields assumed in \simba;  we note that \simba\ reproduces the observed $z=0$ stellar mass -- gas-phase metallicity relation fairly well~\citep{dave_2019}, although this is primarily driven by the oxygen abundance.  Meanwhile, \simba's quenched galaxies are much less sensitive to ionising background and are not obviously in disagreement with the corresponding COS-Halos observations, modulo the large uncertainties in the data, and the fact that most of the absorbers are close to the detection limit.  Larger samples of \MgII absorbers exist around more massive quenched galaxies~\citep[e.g.][]{zahedy_2019}, and we plan to compare to such samples in the future.  Similarly, for \SiIII\ around quenched galaxies, \simba predicts absorption that is within the range of the large uncertainties.  This indicates that \simba\ quenched halos do contain at least some cool absorbing gas, whose origin is an interesting question that we will investigate in future work.

% Carbon
\CIV\ (1548,1550\AA) is more easily traceable in COS-Dwarfs owing to its $z\sim 0$ sample, while at COS-Halos redshifts it moves beyond the COS/G160M grating's coverage. This is an interesting mass regime because for a $10^{9.5} \msolar$ galaxy, the typical virial temperature of its halo is $\sim 10^{5.3}$K, which is near the collisional ionisation peak of \CIV.  Hence \CIV\ can trace both photo-ionised and collisionally ionised gas in COS-Dwarfs, since 
as Figure \ref{fig:mass_budget_omega} shows, both star forming and quenched galaxies have substantial gas in the hot phase.

Comparing to observations, the equivalent widths of \CIV absorbers around star forming galaxies agree in the central region, but \simba over-predicts them in the outskirts. Around quenched galaxies, \simba over-predicts \CIV absorption at a similar level, though the statistical significance is lower. Assuming a different ionising background does not result in any significant change in absorption, suggesting that the ionising background has less impact at these lower masses for \CIV. Generally, \simba doesn't reproduce as strong a radial gradient in EWs as seen in the COS-Dwarfs data. This is also mildly seen in the radial trend for \HI. This could indicate that the metals in \simba are being distributed too far via outflows around low-mass galaxies, although this issue is not evident around the higher-mass COS-Dwarfs galaxies. Alternatively, it could be that the ionisation conditions are incorrect in the outskirts of \simba dwarf halos.  

% Oxygen
Turning to \OVI\ in COS-Halos (lower right), assuming either an \citetalias{faucher-giguere_2020} or \citetalias{haardt_2012} background \simba slightly over-predicts \OVI\ absorption around star forming galaxies by a factor of $\sim 2$. Around quenched galaxies, \OVI\ absorption is in reasonable agreement within the uncertainties, and reflects the dichotomy in \OVI\ absorption around quenched vs. star-forming galaxies owing to oxygen moving into higher ionisation states in the hot gas around quenched systems~\citep{oppenheimer_2020}. Strong \OVI\ absorption in the CGM has been associated with high star formation activity \citep[e.g.][]{tumlinson_2011,zahid_2012b, suresh_2017} since oxygen is produced in Type~II SNe events, but simulations suggest that the oxygen in today's CGM was deposited many Gyr ago~\citep{ford_2014}. Interestingly, previous studies of metal-line absorption in simulations \Romeel{ \citep[e.g.][]{hummels_2013,ford_2016,liang_2016,oppenheimer_2016,gutcke_2017,suresh_2017} with the exception of IllustrisTNG~\citep{nelson_2018b} have} found an under-prediction of \OVI\ around star forming systems by a factor of a few -- \simba is much closer in agreement to the observations on this front. The increase seen in \simba owes in part to our choice of ionising background -- assuming instead the \citetalias{haardt_2001} background as was done in \citet{ford_2016} results in an order of magnitude lower level of \OVI\ absorption, as seen by the dotted line.
\Romeel{This highlights that comparisons of \ion{O}{vi} absorption between models can be sensitive to the ionising background assumed, and without better empirical constraints on the strength of the extreme UV background, it is difficult to make robust comparisons to data.
In contrast,} as with the lower ions, \OVI absorption around quenched systems is much less sensitive to ionising background. Another factor may be that \simba's outflows are ejected 30\% in SN-heated gas, while the \citet{ford_2016} simulations ejected winds solely in a cool phase.

Overall, \Romeel{depending on tracer, distance, and assumed UVB, \simba\ either reasonably reproduces the observed \ion{H}{i} and metal line absorption in the CGM, under-predicts the median EW by up to an order magnitude, or over-predicts EW by a factor of a few.  Particularly interesting discrepancies include the} deficit of \HI\ absorption around quenched galaxies, an under-prediction of \SiIII, and a too-weak radial gradient in \CIV.  The choice of assumed ionising background is an important uncertainty in comparisons to observations, although the results for more modern determinations (\citetalias{haardt_2012} and \citetalias{faucher-giguere_2020}) are more similar.

\subsubsection{Covering fractions}

\begin{figure*}
	\includegraphics[width=0.95\textwidth]{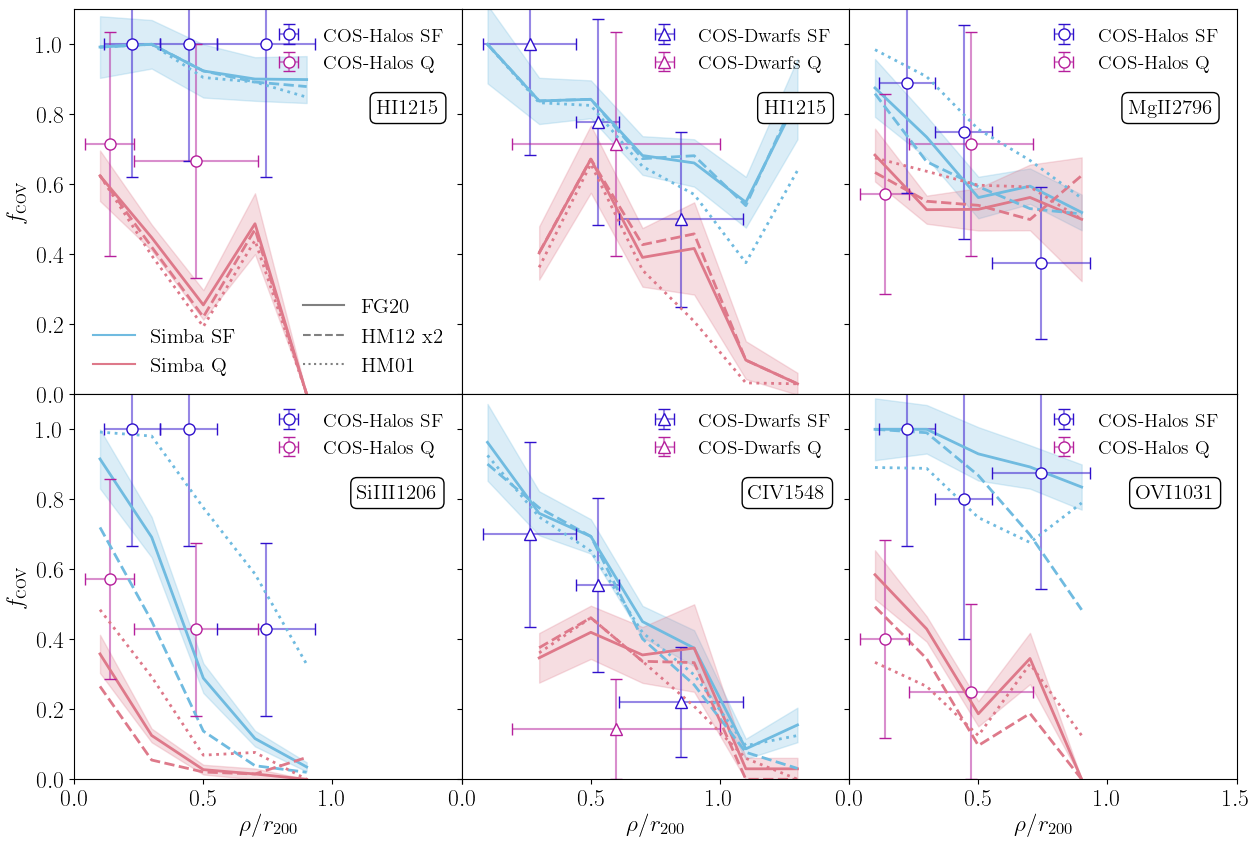}
	\vskip-0.1in
    \caption{\fcov for \HI and selected metal lines (as indicated in each panel) against $r_{200}$-scaled impact parameter for the COS-Halos and COS-Dwarfs galaxy samples, using the \protect\citetalias{faucher-giguere_2020} (solid), \protect\citetalias{haardt_2012} (dashed) and \protect\citetalias{haardt_2001} (dotted) ionising backgrounds. Light blue and light pink lines represent star forming and quenched galaxies in the \simba sample. The shaded regions indicate typical Poisson errors. Dark blue and magenta points represent star forming and quenched galaxies in the COS samples (circles for COS-Halos, triangles for COS-Dwarfs); horizontal error bars indicate the width of the bins and vertical error bars are Poisson errors. With a few exceptions, \simba reasonably reproduces observed \fcov for most species, particularly the dichotomy between star forming and quenched systems seen in \OVI.
    } 
    \label{fig:rho_cfrac}
\end{figure*}

Figure \ref{fig:rho_cfrac} shows \fcov for each ion against $r_{200}$-scaled impact parameter for the COS-Halos and COS-Dwarfs galaxy samples, using the \citetalias{faucher-giguere_2020} (solid), \citetalias{haardt_2012} (dashed) and \citetalias{haardt_2001} (dotted) ionising backgrounds. \simba star forming and quenched galaxies are represented by light blue and light pink lines, respectively; shaded regions show the typical Poisson errors. Star forming and quenched COS-Halos and COS-Dwarfs observations are represented by dark blue and magenta points, respectively. Horizontal error bars on the observations indicate the width of the bins, and vertical error bars are Poisson uncertainties. \Romeel{We note that we have applied the same detection thresholds (listed in Table~\ref{table:thresh}) to the observations sample to ensure a fair comparison.}

In the observations and in \simba, the star forming galaxies show a higher \fcov than quenched galaxies.  Also, in general the \simba galaxies show a radial decrease in \fcov for both the COS-Halos and COS-Dwarfs samples, but this trend is less clear in the data given the errorbars.  These trends broadly mimic those seen for the observed EWs.

For \HI\ around star-forming systems, \simba shows near unity covering fraction around the larger galaxies in COS-Halos, but a strong negative gradient in \fcov\ for the dwarf sample. These trends impressively follow those seen in the data, suggesting that \simba reasonably represents the trend in cool halo gas versus stellar mass.  In contrast, the quenched galaxies show more \HI\ in the outskirts than seen in \simba.  The extremely high \fcov\ at large radii in COS-Halos is curious and highly discrepant from \simba, though it may be a product of small number statistics.  

The trends for the metal ions broadly follow those seen for the EWs. \MgII is in good agreement with observations, \mycolor{with a higher covering fraction around star forming galaxies than quenched galaxies as also observed in larger samples of \MgII absorbers \citep[e.g.][]{lan_2014, lan_2020}.} \SiIII around star-forming galaxies in \simba is low  by $\sim$30\% for \citetalias{faucher-giguere_2020} and \citetalias{haardt_2012}, but in good agreement if \citetalias{haardt_2001} is assumed. The discrepancy in \SiIII around quenched galaxies is more pronounced.  \fcov of \CIV is $\sim$20\% too low at all impact parameters. Perhaps the most impressive agreement is for \OVI, for which a large dichotomy between the star-forming and quenched populations is seen, and nicely reproduced in \simba; we note that this arises without any specific tuning, although there is minor variation in this owing to the choice of ionising background.  With modest exceptions, \simba generally reproduces the observed covering fractions around these galaxy populations.

\subsubsection{Path absorption}

\begin{figure*}
	\includegraphics[width=0.95\textwidth]{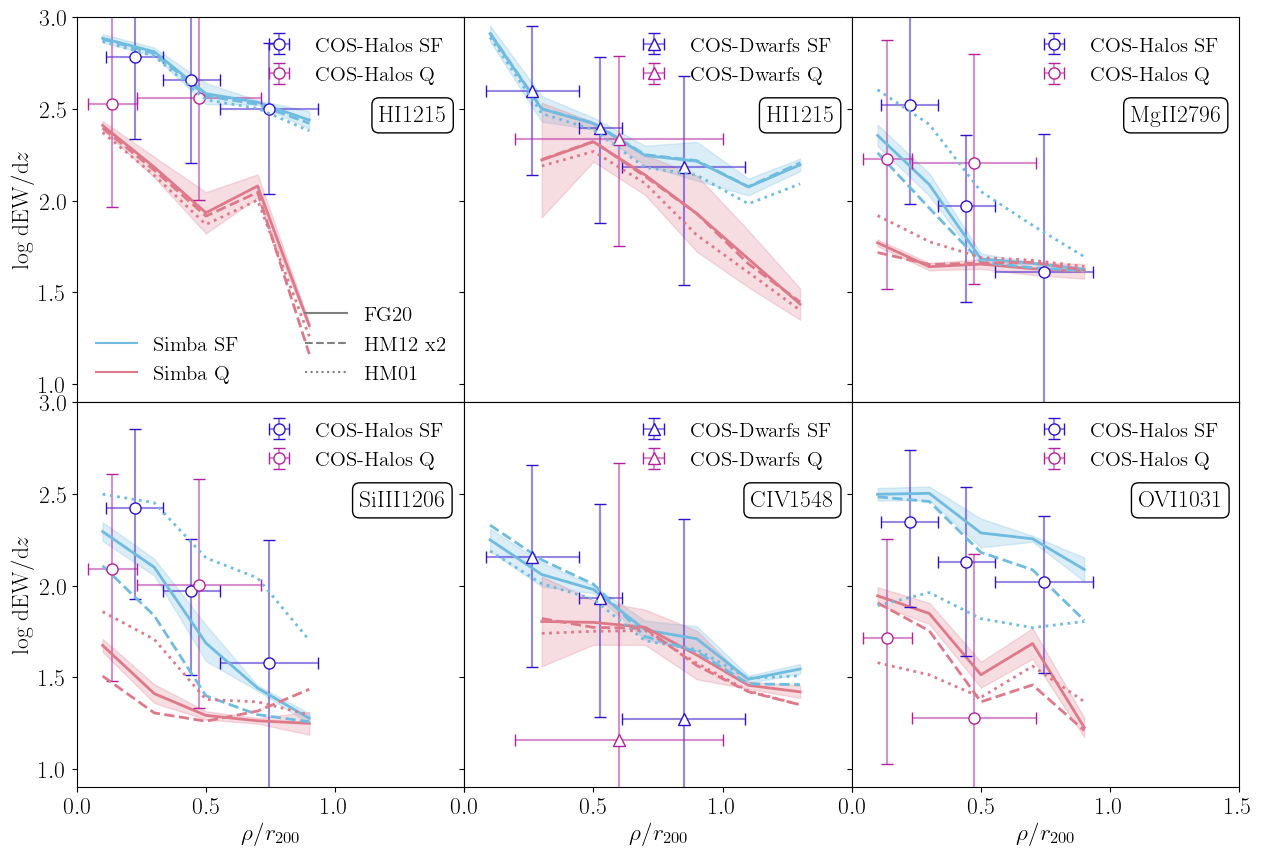}
	\vskip-0.1in
    \caption{Total path absorption of \HI and selected metal lines (as indicated in each panel) against $r_{200}$-scaled impact parameter for the COS-Halos and COS-Dwarfs galaxy samples, using the \protect\citetalias{faucher-giguere_2020} (solid), \protect\citetalias{haardt_2012} (dashed) and \protect\citetalias{haardt_2001} (dotted) ionising backgrounds. Light blue and light pink lines represent star forming and quenched galaxies in the \simba sample; shaded regions show typical cosmic variance uncertainties. Dark blue and magenta points represent star forming and quenched galaxies in the COS samples (circles for COS-Halos, triangles for COS-Dwarfs); horizontal error bars indicate the width of the bins while vertical error bars are the standard deviations from computing path absorption individually for each galaxy. \simba closely reproduces total path absorption of \HI around star forming galaxies; for metal lines in the COS-Halos sample, our results are sensitive to ionising background. } 
    \label{fig:path_abs}
\end{figure*}

Figure \ref{fig:path_abs} shows the path absorption $dEW/dz$ for each ion against $r_{200}$-scaled impact parameter for the COS-Halos and COS-Dwarfs galaxy samples, using the \citetalias{faucher-giguere_2020} (solid), \citetalias{haardt_2012} (dashed) and \citetalias{haardt_2001} (dotted) ionising backgrounds. As in the previous figures, \simba star forming and quenched galaxies are represented by light blue and light pink lines respectively; shaded regions show the typical cosmic variance uncertainties. 

Dark blue and magenta points represent the COS-Dwarfs and COS-Halos observations; horizontal error bars indicate the width of the bins and vertical error bars are the standard variation within each bin. \Romeel{For the data, we compute} these values \Romeel{by summing the equivalent widths of all the lines} above the detection thresholds as indicated in Figure \ref{fig:rho_ew}\Romeel{, and then dividing by the $\Delta z$.  We cannot fully mimic the procedure that we do in the simulations of adding up all the equivalent widths directly from the spectra because} \response{this would involve reducing the spectra and associating absorption features with specific ions, which is an involved process that is beyond the scope of this work}. Nonetheless, at least in \simba, the path absorption around star-forming systems is dominated by lines above these detection thresholds, so it is reasonable to compare these results directly to observations \Romeel{since the uncounted equivalent width below the detection limit in the data will not dominate}.  In contrast, for the quenched galaxies, this is typically not the case; \response{hence this comparison should be viewed with some caution, nonetheless we show it as the red lines}. Our predicted path absorption could be compared to data of any quality, so it provides a more general prediction for future spectral stacking experiments to quantify weak CGM absorption.

Overall, the results for total path absorption closely resemble that of equivalent width. As seen previously with equivalent width and covering fractions, total path absorption is higher in star forming galaxies and decreases as a function of impact parameter. \simba closely reproduces the total path absorption of \HI around star forming \lstar galaxies and broadly produces metal line total path absorption that generally agrees with the COS-Halos observations, however \MgII and \SiIII (\OVI) are slightly under (over) predicted. \simba also closely reproduces the total path absorption of \HI and \CIV around sub-\lstar star forming galaxies, with slight over-prediction of \CIV in the outskirts. Thus in general, the path absorption tells a similar story to the other measures, although with improved statistics it could potentially highlight important differences in the weak absorber population.

\subsection{Summary of observational comparisons}

\simba broadly reproduces the COS-Halos and COS-Dwarfs observations of \HI and metal CGM absorbers, but there are some significant discrepancies as well.  A notable success is that the simulations reproduce the near-unity \HI covering factors around star forming \lstar galaxies, together with the strong radial gradient around dwarfs.  In contrast, \simba under-produces \HI around quenched galaxies.  \citet{van_de_voort_2019} and \citet{hummels_2019} have pointed out \Romeel{using CGM-focused zoom simulations} that the \HI content of \lstar halos, which typically contain substantial hot gas, is quite sensitive to numerical resolution \Romeel{\citep[though see][who found little sensitivity due to the already high resolution]{suresh_2019}}. It could be that \simba simply lacks the required resolution to model condensations of \HI within hot halos, even while reproducing the \HI in halos that have substantial cool gas.

For metal lines, the low ions \MgII and \SiIII are somewhat under-produced around star-forming galaxies.  This might owe to these ions tracing denser CGM gas for which our predictions might be particularly sensitive to resolution effects. However, in Appendix \ref{sec:convergence} we show that the predictions are not strongly sensitive to resolution, for the range of resolutions we can probe among our runs. Meanwhile, the low ion predictions are quite sensitive to the ionising background; switching to HM01 yields quite good agreement, consistent with \citet{ford_2016}.  Hence it is difficult to interpret this disagreement without independent constraints on the strength and shape of the ionising background.

\OVI present a significant success for \simba.  Previous works highlighted the challenge for cosmological models to produce enough \OVI absorption around star-forming galaxies, but this is partially mitigated in \simba even when using the same ionising background (HM01), and disappears completely using a more recent background (HM12 or FG20). Moreover, \simba nicely reproduces the dichotomy in \OVI absorption between star forming and quenched systems, as also seen in IllustrisTNG and EAGLE~\citep{nelson_2018b,oppenheimer_2020}. This provides circumstantial evidence that sufficient \OVI arises from a volume filling warm-hot phase~\citep{ford_2013} rather than interfaces of cold clouds in a hot medium~\citep{heckman_2002} that would be unresolved in \simba.

\Romeel{Numerical convergence is nonetheless an important systematic in these comparisons.  In Appendix~B we examine the convergence with respect to both simulation resolution and volume in the COS-Halos and COS-Dwarfs comparisons above.  In general, higher resolution produces more absorption overall, as does smaller volume; the latter owes to less high-mass objects with jet-mode AGN feedback heating.  These variations are at the $\sim 0.1-0.2$~dex level for most ions, though for \CIV in the COS-Dwarfs sample it is higher.  These systematics, comparable to those from varying the ionising background, must be kept in mind when assessing (dis)agreement with the observations.
}

These results encouragingly suggest that \simba is generally populating the CGM with cool gas and metals in accord with absorption line observations, but also highlight some of the challenges in interpreting such observations given current knowledge.  One way forward is to attempt to connect CGM physical and absorption line properties with feedback processes, to understand how the CGM provides direct constraints on baryon cycling.  This is what we do next.

\section{Dependence on feedback models}\label{sec:feedback}

Feedback plays a critical role in setting the properties of galaxies~\citep{somerville_2015,Naab_2017}, but less is understood about how such processes impact the CGM.  
To study the impact of feedback on the CGM, we make use of the following suite of feedback variant \simba simulations: No-Xray (no X-ray AGN feedback); No-jet (no jet-mode or X-ray AGN feedback); No-AGN (no jet-mode or X-ray or radiative AGN feedback) ; and No-feedback (No AGN or SF driven feedback). These are run in $50\hmpc$ volumes with $512^3$ gas elements and dark matter particles, each with identical initial conditions. We note that these variants fail to reproduce key galaxy observations~\citep{dave_2019} -- for instance without jet feedback we get essentially no quenched galaxies -- but instead are intended as numerical experiments to connect specific feedback processes to their impact on the CGM.

By comparing between these variants, we can isolate how SF feedback and \simba's AGN feedback modes impact the physical and observable properties of the CGM.
We first study the composition of the CGM in the AGN variant volumes by focusing on mass fractions and metallicities of the halo components for each galaxy.  We then examine how these variants impact absorber statistics, in particular their path absorption.

\subsection{Halo baryon fractions}

\begin{figure*}
	\includegraphics[width=0.95\textwidth]{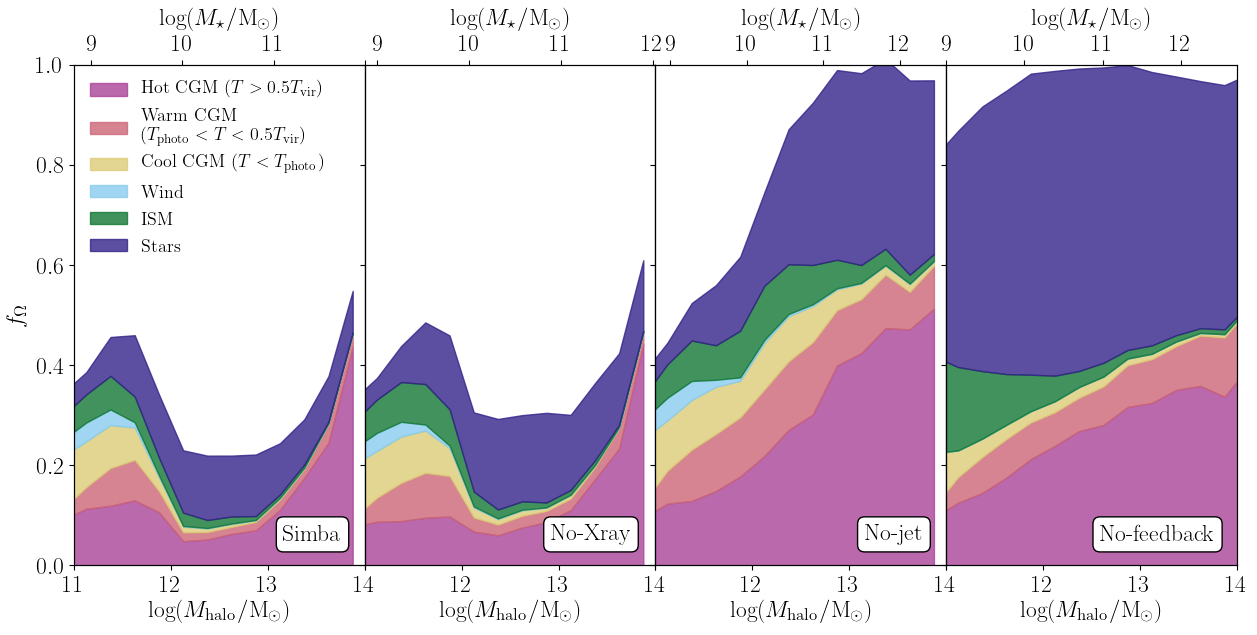}
	\vskip-0.1in
    \caption{Median mass fraction in each halo component relative to $M_{\Omega b}$ as a function of total halo mass for the AGN feedback variant simulations (from left to write: full \simba model, no X-ray AGN feedback, no jet-mode or X-ray AGN feedback, and no AGN or SN feedback). The colour coding of each component is the same as in Figure~\ref{fig:mass_budget_omega}. We omit the model with SN feedback and no AGN feedback from this plot since there is little difference between this and the no-jet model. Stellar feedback reduces baryon fraction at the low \mstar end, while jet-mode AGN feedback heats CGM and ISM gas and strongly suppresses baryon fraction at the high \mhalo end.} 
    \label{fig:winds_omega_frac}
\end{figure*}

Figure \ref{fig:winds_omega_frac} shows the median mass fraction $f_\Omega$ in each component \Romeel{versus halo mass}, scaled by the baryon mass expected if the halo has retained it cosmic share of baryons (i.e. $f_\Omega\equiv M_{\rm halo}\frac{\Omega_b}{\Omega_M}$) for each of the AGN feedback variant simulations. We plot points only where there are $>$10 galaxies in a halo mass bin, and show results for the total galaxy population as there are too few quenched galaxies in the absence of jet-mode feedback. \Romeel{We compare at fixed halo (not stellar) mass in this case because the different variants have quite different efficiencies of star formation, while the halo masses are fairly consistent among these runs (Sorini et al., in prep.).}  The $50\hmpc$ volume reproduces the halo baryon fractions seen previously in the $100\hmpc$ volume (\S\ref{sec:mass_budget}), except that the smaller volume does not probe the highest halo masses; \Romeel{this shows good volume convergence for the halo baryon fractions in the various phases.}

Starting with the no-feedback run in the rightmost panel, we see that in the absence of feedback, all but the smallest halos retain close to their cosmic share of baryons.  The stellar fractions are very high, demonstrating overcooling~\citep{dave_2001,balogh_2001}.  While all halos have virial-temperature gas, the fraction of it increases steadily with halo mass.  These establish the trends predicted without any baryon cycling.

Introducing SF feedback (i.e. the No-jet model) results in a significant suppression of baryons in low-mass halos. This follows the trend assumed in \simba that the mass loading factor in outflows scales inversely with stellar mass, and that the smaller potential wells enable easier escape from low-mass halos.  Small halos are less able to retain their outflows, creating a baryon deficit.

The most dramatic change occurs when we turn on AGN jet feedback (No-Xray), causing a strong reduction in the halo baryon fraction around all but the smallest galaxies. This demonstrates that the jet-mode of our AGN model is responsible for the reduction in halo baryon fraction.  In contrast, by comparing panels it is clear that that X-ray and radiative AGN feedback have comparatively little effect on the overall baryon fraction contained in halos. The reduction in baryon fraction for galaxies with $\mstar \ga 10^{10}\msolar$ occurs when jet feedback begins turning on, and becomes stronger around higher mass galaxies. In the extreme case, at $\mstar \sim 10^{11.5}\msolar$ the fraction decreases from unity to 0.4 in the presence of jet mode feedback. 

Jet feedback also has the most substantive effect on the CGM phases. It heats the halo gas, decreasing the relative fractions of ISM and cool and warm CGM gas, and decreasing the relative fraction in stars at the high mass end due to the overall decrease in star formation in the most massive galaxies. Adding X-ray AGN feedback to get the full \simba physics, we see only modest changes in the CGM mass budgets.  The most dramatic appears to be the increase in hot gas fraction in the most massive halos, owing to X-ray feedback adding heat directly to surrounding gas if it is not star-forming; however, this may be subject to small number statistics in these most massive systems.

\subsection{CGM metallicities}

\begin{figure*}
	\includegraphics[width=0.95\textwidth]{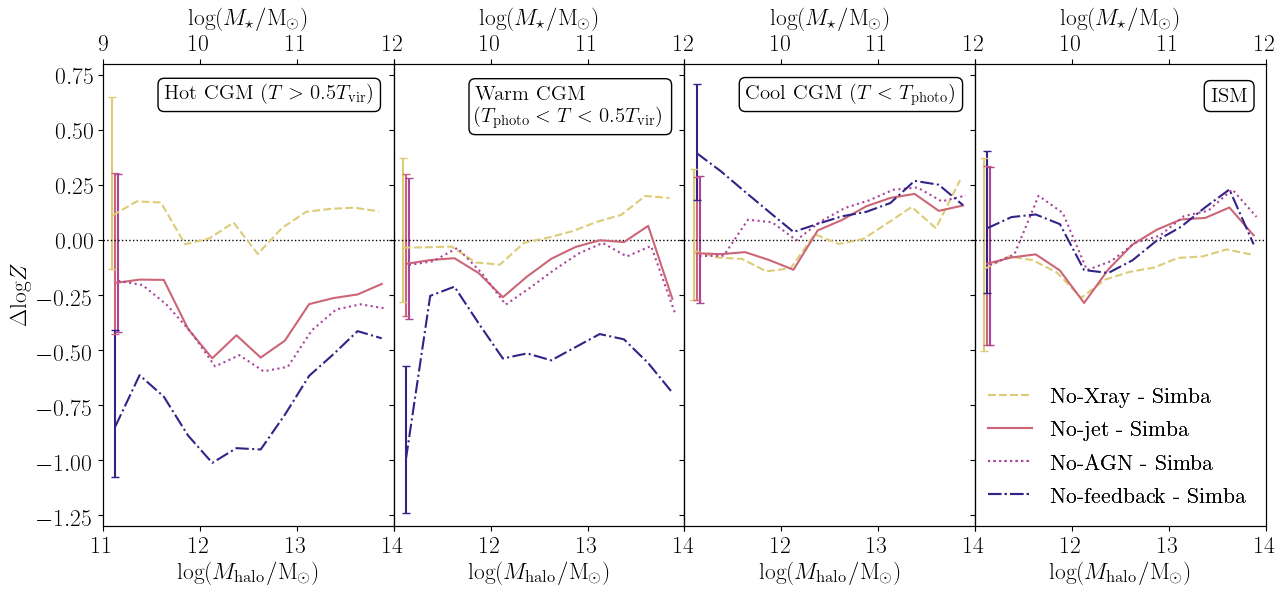}
	\vskip-0.1in
    \caption{Difference in log metallicity between the fiducial \simba run and the variant without X-ray heating (yellow dashed line), without X-ray heating or AGN jets (red solid line), with stellar feedback only (magenta dotted line), and with no feedback prescription at all (dark blue dot-dashed line). Each panel refers to a different gas phase, as reported inside the plots. The error bars indicate the 25$^{\rm th}$ and 75$^{\rm th}$ percentiles of the data; for simplicity we show one representative error bar per line. Very low $\Delta {\rm log} Z$ in the warm gas component are plotted at -2. Points are slightly offset horizontally for easier viewing. Stellar feedback increases the metallicity of the warm and hot CGM components, while there is little impact on the cool CGM and ISM gas.
    } 
    \label{fig:winds_gas_metallicities}
\end{figure*}

Next we present the effect of feedback on the metallicity of the halo gas components.  Figure \ref{fig:winds_gas_metallicities} shows the difference in median log metallicity between \simba and each of the AGN feedback variant simulations, as a function of halo mass. A value above (below) zero indicates that galaxies in the AGN variant simulation have higher (lower) metallicity than in the main \simba volume. We show one representative error bar per line; the error bars indicate the 25th and 75th percentiles of the data. As before we plot points only where there are $>$10 galaxies in a halo mass bin, and show results for the total galaxy population. 

In the no feedback simulation metals are primarily locked into stars, resulting in a high stellar metallicity and low metallicity in the hot and warm phase CGM (lower than in the full \simba model by 1.5-0.5 dex, from the low $\mstar$ end to high $\mstar$ end.). Within the uncertainties, the metallicities of the cool CGM and the ISM are not affected by the AGN and SF feedback model.

Turning on SF feedback raises the metallicity of the warm and hot CGM, seen in the increase in $\Delta {\rm log} Z$ between the No-feedback and No-AGN models. This is especially true at the low mass end where SF feedback is most effective. Despite 70\% of SN wind particles being ejected cold at $T \sim 10^3$K, the lack of a change in metallicity in the cool CGM phase indicates that these ejected metals do not remain cold. Instead the increase seen in the warm and hot components indicates that the metal-enriched gas is heated towards the halo virial temperature.

Radiative-mode AGN feedback has little impact on the hot CGM. However, introducing jet feedback raises this metallicity by 0.5 dex, possibly because metal-rich gas that was previously in the cool or warm phases is heated by the jets to the virial temperature.

\subsection{CGM absorption}

Finally, we examine the impact of feedback on CGM absorption, by repeating the COS-Halos and COS-Dwarfs comparison using the feedback variant simulations. We use the same galaxy selection procedure as section \ref{sec:galaxy_selection} to select suitable galaxies from the $50\hmpc$ \simba volume, \mycolor{in this case selecting 4 \simba galaxies per COS galaxy to account for reduced volume}. We then identify matching halos from the AGN variant simulations to probe the effect of AGN feedback on specific halos, where matching halos are defined as having the most number of dark matter particles in common. This is possible because all the feedback variants are run from identical initial conditions, so the dark matter particles have the same IDs. We ignore some high mass galaxies from the COS-Halos sample that have no analogs in the small $50\hmpc$ \simba volume.

\begin{figure*}
	\includegraphics[width=0.95\textwidth]{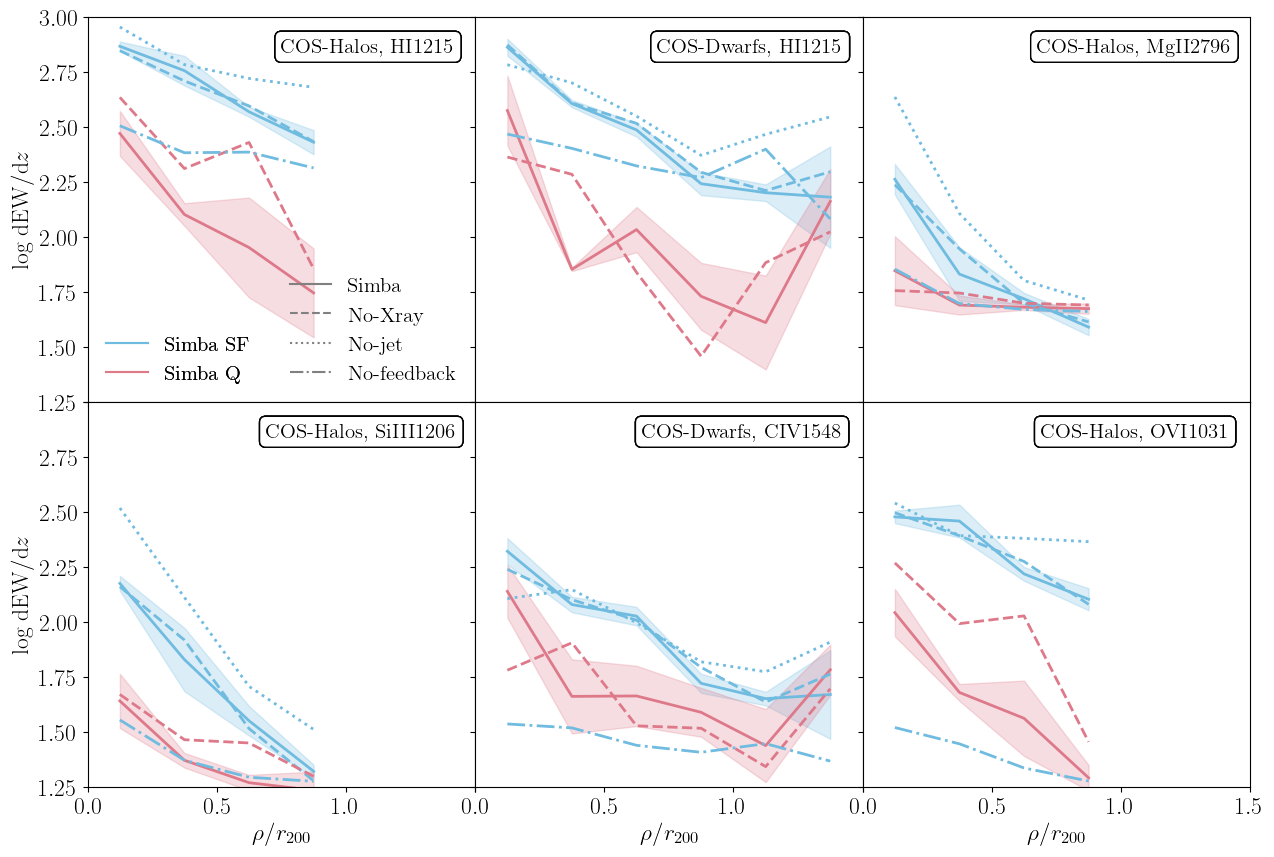}
	\vskip-0.1in
    \caption{Total path absorption for different AGN models (solid: full \simba; dashed: No-Xray; dotted: No-jet; dot-dashed: No-feedback) for \HI and selected metal lines (as indicated in each panel), against $r_{200}$-scaled impact parameter in the COS-Halos and COS-Dwarfs \simba galaxy samples. We assume an \protect\citetalias{faucher-giguere_2020} ionising background. Light blue and light pink lines represent star forming and quenched galaxies, respectively; shaded areas show the cosmic variance uncertainties. Stellar feedback enriches the CGM leading to increased absorption, while jet-mode AGN feedback heats the CGM and leads to decreased absorption around \lstar galaxies.
    } 
    \label{fig:winds_path_abs}
\end{figure*}

Figure \ref{fig:winds_path_abs} shows the total path absorption against $r_{200}$-scaled impact parameter in the COS-Halos and COS-Dwarfs \simba galaxy samples, using the full \simba model (solid), the No-Xray model (dashed), the No-jet model (dotted) and the No-feedback model (dot-dashed). We assume an \citetalias{faucher-giguere_2020} ionising background. Light blue and light pink lines represent star forming and quenched galaxies; error bars show the cosmic variance uncertainties. The quenched galaxies in the No-jet and No-feedback models are not shown due to small number statistics, since jet feedback is the dominant quenching mechanism in \simba. We also choose not to show the No-AGN results as we find these absorption statistics are similar to that of the No-jet simulation, i.e. radiative-mode feedback has little effect on the CGM in absorption in our simulations.

In the absence of SF driven feedback to enrich the CGM, we see reduced CGM absorption for all ions, particularly for \SiIII and \OVI. This is not due to a change in the multiphase structure of the CGM, since Figure \ref{fig:winds_omega_frac} shows that SF feedback does not significantly change the temperature of the CGM. Instead SF feedback enriches the CGM by spreading metals away from central galaxies across a range of ionisation potentials.  Hence SF feedback is primarily responsible for enriching the CGM.

In the \lstar galaxy COS-Halos sample, absorption around star forming galaxies increases in the absence of X-ray and jet feedback. Turning on jet feedback leads to overall heating of galaxy halos and the IGM \citep{christiansen_2020}, resulting in reduced absorption for all ion species considered here. The decrease is largest for low ions, which trace cool gas, and at the outskirts of the halos, where gas is additionally heated due to jet feedback from neighbouring galaxies. The decrease in absorption around \lstar galaxies from turning on X-ray feedback is due to the direct heating of non-ISM gas, which scales with X-ray flux. X-ray feedback has no effect on the sub-\lstar galaxies in the COS-Dwarfs sample, which typically do not have massive black holes with the $\fedd < 0.2$ required for jet and X-ray feedback. Jet feedback only reduces absorption further away from these galaxies at $\gtrsim r_{200}$, where the IGM is heated by jet feedback from nearby galaxies.  These results show that SF feedback and jet-mode AGN feedback have the largest impact on CGM absorption properties, the former providing the metals, and the latter important for setting the gas phase. Thus, observational surveys such as COS-Halos and COS-Dwarfs may provide constraints on the amount of metals transported in stellar winds, and on AGN jet velocity.

\section{Conclusions}\label{sec:conclusions}

We have examined the physical and observable properties of CGM gas around galaxies of a wide range in mass and specific SFR in the \simba cosmological hydrodynamic simulation at low redshifts.  We examine halo baryon and metal contents as a function of mass, and conduct a detailed comparison to \HI and metal absorption line observations in the COS-Halos and COS-Dwarfs databases.  Finally, we examine the impact of various star formation and AGN feedback modules in \simba on CGM properties.  Our main results are as follows:

\begin{itemize}
    \item Baryon fractions in halos across the entire mass range probed are well below the cosmic fraction.  This indicates that missing halo baryons are truly \mycolor{redistributed beyond $r_{200}$}, not just undetected.  The CGM of \lstar\ galaxies has largely been evacuated, down to as little as $\sim 10\%$ of the global baryon fraction in CGM gas.  Star-forming systems, even \mycolor{at the high-mass end,} tend to have a CGM with a comparable mix of cool $\sim 10^4$K gas and gas near the virial temperature, while for quenched systems, \mycolor{even at the low-mass end,} the CGM \mycolor{gas is near the virial temperature}.  Thus the distinction between galaxies having hot vs. multi-phase CGM is not so much of a division in mass but rather one in sSFR.
    \item For star-forming systems, metals are increasingly locked into galactic components (stars, dust, ISM) towards higher $\mstar$, reflecting the increased efficiency of star-formation up to the highest masses.  For the CGM, metals are distributed among the phases, with the hotter phases having a larger share at higher $\mstar$.  In quenched systems, in contrast, CGM metals are mostly hot, reflecting the overall mass phase budget, and there is no trend with mass except a mild increase at high masses where the star formation efficiency drops.
    \item CGM metallicities show strong trends with temperature, as the virial-temperature CGM gas has $\sim0.5-1$ lower metallicity than cooler phases.  The cool and warm CGM, in contrast, have metallicities close to the ISM, which itself generally tracks stellar metallicities.  The lower hot phase metallicity likely reflects \mycolor{less metal enriched halo inflow} that is virialised, mixing with metals deposited via outflows. 
    \item Absorption line statistics around a matched sample of \simba galaxies are broadly in agreement with the COS observations.  Equivalent width, covering fraction and total path absorption decrease with increasing $r_{200}$-scaled impact parameter for each ion species. CGM absorption is higher around star forming galaxies than around quenched galaxies; the difference is smaller around sub-\lstar galaxies, suggesting lower-mass galaxies are quenched via a different mechanism.

    \item \simba predictions of \HI equivalent width and total path absorption around star forming galaxies are in good agreement with COS-Halos and COS-Dwarfs observations, while covering fractions are under-predicted around \lstar star forming galaxies and too extended around sub-\lstar galaxies. \simba under-predicts \HI EW and \fcov around quenched galaxies.
    \item Metal line absorption around \lstar galaxies is sensitive to choice of photoionising background. When assuming a \citetalias{faucher-giguere_2020} background, \simba under-predicts EW, \fcov and total path absorption of low ions \MgII and \SiIII. Assuming instead a \citetalias{haardt_2001} brings the predictions into closer agreement with COS-Halos observations. Around quenched galaxies, \simba does not reproduce as strong a radial gradient of low ion absorption seen in COS-Halos.  
    \item \simba reproduces observed \OVI absorption separately in star-forming and quenched systems, when employing more recent ionising background determinations.  Reproducing the observed dichotomy is thus an important success of \simba's model for halo gas heating associated with galaxy quenching, and supports the scenario that \OVI mostly arises from a smoother volume-filling warm phase~\citep{ford_2013} that gets ionised to higher states around quenched galaxies~\citep{oppenheimer_2016}.
    \item Around sub-\lstar galaxies, EW, \fcov and total path absorption of \CIV around star forming galaxies are in good agreement with COS-Dwarfs observations in the central region, but \simba over-predicts in the outskirts. Around quenched galaxies, \simba over-predicts \CIV at all impact parameters.
    \item Feedback strongly lowers the mass fraction in the CGM.  With no explicit feedback, halos contain roughly their cosmic share of baryons.  Star-formation feedback increasingly evacuates the CGM below \lstar, while jet AGN feedback has a dramatic impact at $\ga$ \lstar. Radiative and X-ray AGN feedback have minimal effects.  Thus in \simba, the same jets that enact galaxy quenching are largely responsible for setting the CGM contents around non-dwarf galaxies. 
    \item Feedback has little impact on the metallicity of the ISM and cool CGM at a given $\mstar$.  For warmer phases, in contrast, star formation feedback has the greatest impact.  This indicates that hot halo gas is primarily enriched via galactic winds, likely at earlier epochs~\citep{oppenheimer_2012,ford_2014}.  The AGN jets have a noticeable impact as well, mildly increasing the hot CGM metallicity.
    \item The above trends among feedback variants are reflected in CGM absorption, with the enrichment of the CGM by star formation driven feedback having the most dramatic impact at raising absorption in all ions including \HI.  AGN jet feedback sets the gas phase of the CGM and tends to counteract this to reduce absorption in the lower ions, but has little impact on \OVI, and essentially no impact on the COS-Dwarfs results since jets are not active in such galaxies. 
\end{itemize}

Our results highlight the complex interplay between galaxy formation, large-scale structure evolution, and feedback processes in setting the physical and observable properties of the CGM.  \simba provides a self-consistent cosmological description of the CGM that is observationally concordant within current systematics \Romeel{at the $\la 2\sigma$ level for a given ionising background}, while predicting interesting results such as a dramatic evacuation of CGM gas, a strongly lower metallicity in the hot CGM phase, and much different trends around star-forming and quenched galaxies regardless of mass, a trend also seen in IllustrisTNG \citep{nelson_2018b}.

Current CGM absorption measures at low-$z$ do not tightly constrain models, owing not only to small numbers but also systematic uncertainties such as a significant dependence of metal absorption on the shape of the ionising background.  In our case, we have only considered spatially-uniform backgrounds, but low ions arising close to galaxies could further be impacted by the escape of ionising photons from the host galaxy.  On the modeling side, numerics remain an important systematic, as idealised CGM models suggest extremely small scale structures that are unresolvable in any cosmological setting, particularly related to how outflows interact with ambient gas~\citep[e.g.][]{fielding_2020a}; new subgrid prescriptions such as Physically Evolved Winds~\citep[PhEW;][]{huang_2020} aim to better account for such effects.  \simba's resolution is hopelessly far from that required to directly model these small-scale interactions, but it remains to be seen how much such interactions drive bulk CGM properties.  With current modeling capabilities, statistically comparing to absorption line surveys spanning a range of galaxy masses and environments as we have done in this work is only possible at cosmological resolutions.  Clearly, much work remains to connect large-scale models such as \simba with small-scale work illuminating the detailed physics of the CGM in order to provide a more robust description of the CGM and its observable properties.

\section*{Acknowledgements}

\mycolor{The authors thank the referee for their helpful comments and suggestions which have improved the paper.} 
We acknowledge helpful discussions with Jacob Christiansen, Shuiyao Huang, Christopher Lovell, Neal Katz, Katarina Kraljic, Benjamin Oppenheimer \& Dylan Robson.
SA thanks Rongmon Bordoloi and Claude-Andr\'e Faucher-Gigu\`ere for sharing their data.
We thank Philip Hopkins for making \gizmo\ public, Robert Thompson for developing \caesar, Horst Foidl, Thorsten Naab and Bernhard Roettgers for developing \pygad, and J. Xavier Prochaska, Nicolas Tejos and Joe Burchett for developing \pyigm.
SA is supported by a Science \& Technology Facilities Council (STFC) studentship through the Scottish Data-Intensive Science Triangle (ScotDIST).
% RD, DS, KSF, BS
RD acknowledges support from the Wolfson Research Merit Award program of the U.K. Royal Society.
DS is supported by the European Research Council, under grant no. 670193.

\simba was run on the DiRAC@Durham facility managed by the Institute for Computational Cosmology on behalf of the STFC DiRAC HPC Facility. The equipment was funded by BEIS (Department for Business, Energy \& Industrial Strategy) capital funding via STFC capital grants ST/P002293/1, ST/R002371/1 and ST/S002502/1, Durham University and STFC operations grant ST/R000832/1. DiRAC is part of the National e-Infrastructure. 

\section*{Data and Software Availability}

The simulation data underlying this article are publicly available\footnote{\url{https://simba.roe.ac.uk}}. The software used in this work is available on Github\footnote{\url{https://github.com/sarahappleby/cgm}} and the derived data will be shared on request to the corresponding author.

%%%%%%%%%%%%%%%%%%%%%%%%%%%%%%%%%%%%%%%%%%%%%%%%%%


\bibliographystyle{mnras}
\bibliography{main} % if your bibtex file is called example.bib

%%%%%%%%%%%%%%%%%%%%%%%%%%%%%%%%%%%%%%%%%%%%%%%%%%

%%%%%%%%%%%%%%%%% APPENDICES %%%%%%%%%%%%%%%%%%%%%

\appendix

\section{Physical impact parameter}\label{sec:kpc_units}

\begin{figure*}
	\includegraphics[width=0.95\textwidth]{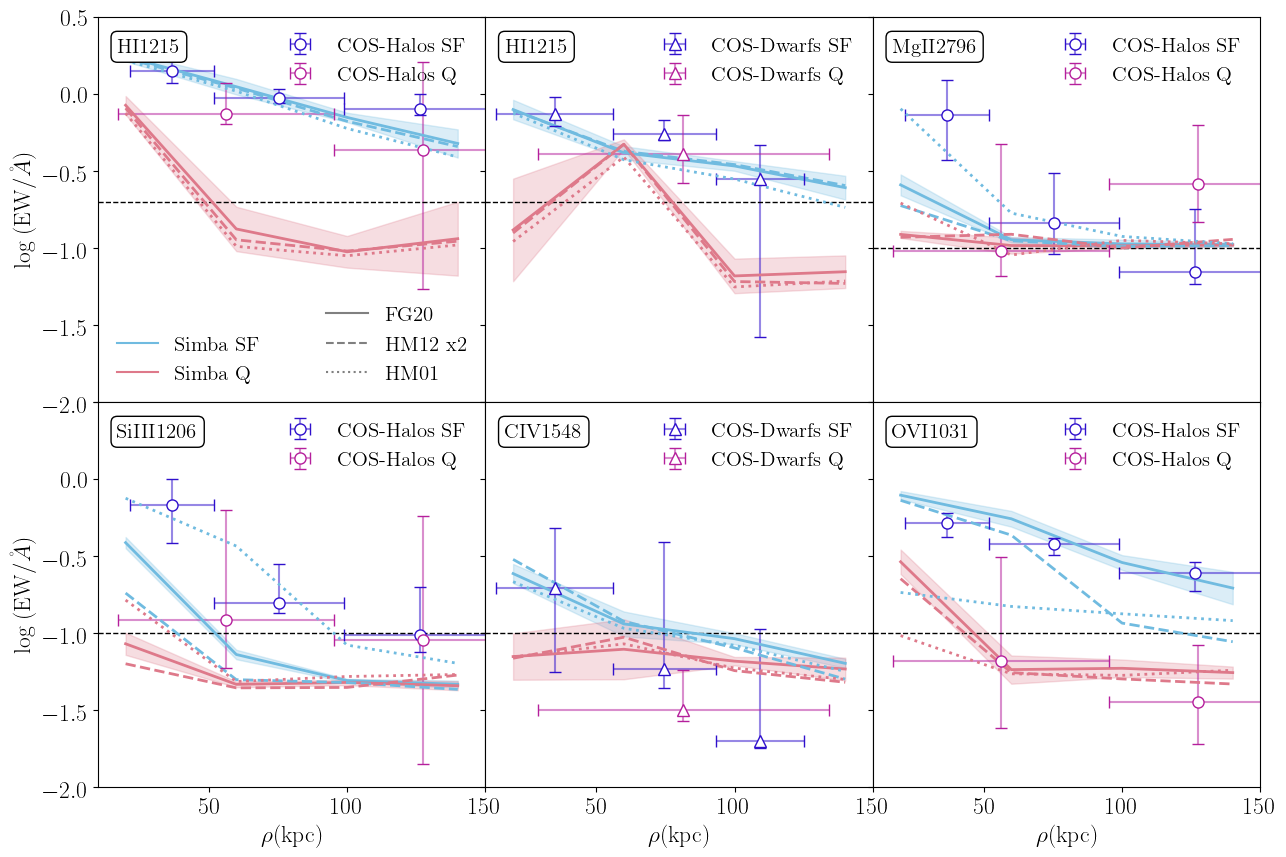}
	\vskip-0.1in
    \caption{Equivalent width of \HI and selected metal lines (as indicated in each panel) against impact parameter for the COS-Halos and COS-Dwarfs galaxy samples (circles and triangles, respectively). Light blue and light pink lines represent star forming and quenched galaxies in the \simba sample, respectively; shaded regions show the cosmic variance uncertainties. Dark blue and magenta points represent star forming and quenched galaxies in the COS samples, respectively; vertical error bars represent the 25$^{\rm th}$ and 75$^{\rm th}$ percentiles of the data, while horizontal error bars indicate the width of the bins. Black horizontal dotted lines indicate the detection threshold of each line under the COS survey conditions.} 
    \label{fig:rho_ew_unscaled}
\end{figure*}

Previous work using the COS-Halos and COS-Dwarfs data have typically displayed spatial absorption dependence in terms of impact parameter in physical kpc units \citep[e.g.][]{werk_2013, tumlinson_2013, bordoloi_2014, ford_2016}. Here we provide the equivalent widths of Figure \ref{fig:rho_ew} as a function of physical impact parameter as these units may be more familiar to readers. Figure \ref{fig:rho_ew_unscaled} shows median equivalent width of selected ions against impact parameter for our COS-Halos and COS-Dwarfs galaxy samples, using the \citetalias{faucher-giguere_2020} (solid), \citetalias{haardt_2012} (dashed) and \citetalias{haardt_2001} (dotted) ionising backgrounds. As before \simba\ star forming and quenched galaxies are represented by light blue and light pink lines, and shaded regions indicate the cosmic variance within eight sub-quadrants over the simulation volume for the \citetalias{faucher-giguere_2020} results. Dark blue and magenta points represent the actual COS-Dwarfs/COS-Halos data; the error bars are the 25th and 75th percentiles of the data. The varying bin widths are chosen such that each bin contains $\sim8$ galaxies.  The detection thresholds for each ion are indicated by horizontal dotted lines. 

Comparison with Figure \ref{fig:rho_ew} demonstrates that aligning the galaxies on a common scale results in stronger radial trends, although the overall trends do not change significantly when using physical kpc units. The \OVI absorption is in slightly better agreement with observations using the physical kpc units, although the difference is within a factor of 2.  This illustrates the level of systematic uncertainty on comparing to observations using different radial measures.

\section{Resolution and volume convergence}\label{sec:convergence}

To assess the impact of numerical limitations on our results, we perform a convergence test using a higher resolution 25 $\hmpc$ volume with the full \simba physics and $512^3$ gas and dark matter particles. The higher resolution run has 8 times the mass resolution of the fiducial volume, however the smaller volume does not properly probe galaxies at the high \mstar end with the most massive black holes seen in the 100 $\hmpc$ box. Hence heating and preventative feedback due to AGN is weaker in smaller volume simulations~\citep{christiansen_2020}. We quantify the volume convergence by examining the 50 $\hmpc$ volume with $512^3$ gas and dark matter particles (the same run as in section \ref{sec:feedback}), and a second 25 $\hmpc$ volume with $256^3$ gas and dark matter particles. These runs have the same mass resolution as the fiducial volume.

We select a sample of galaxies matching the COS-Halos and COS-Dwarfs criteria using the procedure outlined in section \ref{sec:galaxy_selection}. For the 50 $\hmpc/512^3$ volume, we use the same galaxy sample as in section \ref{sec:feedback}. Due to the low volume of the 25 $\hmpc$ runs, there are insufficient galaxies in these simulations to match every COS survey galaxy, particularly at the high mass end. To account for this we select \mycolor{fewer analogous \simba galaxies (4 per COS galaxy)} and omit COS galaxies from our sample that have insufficient analogous \simba galaxies. Additionally, we do not impose the isolation criteria (no other central galaxies with $1\ \hmpc$) for the 25 $\hmpc$ samples.

\begin{figure*}
	\includegraphics[width=0.95\textwidth]{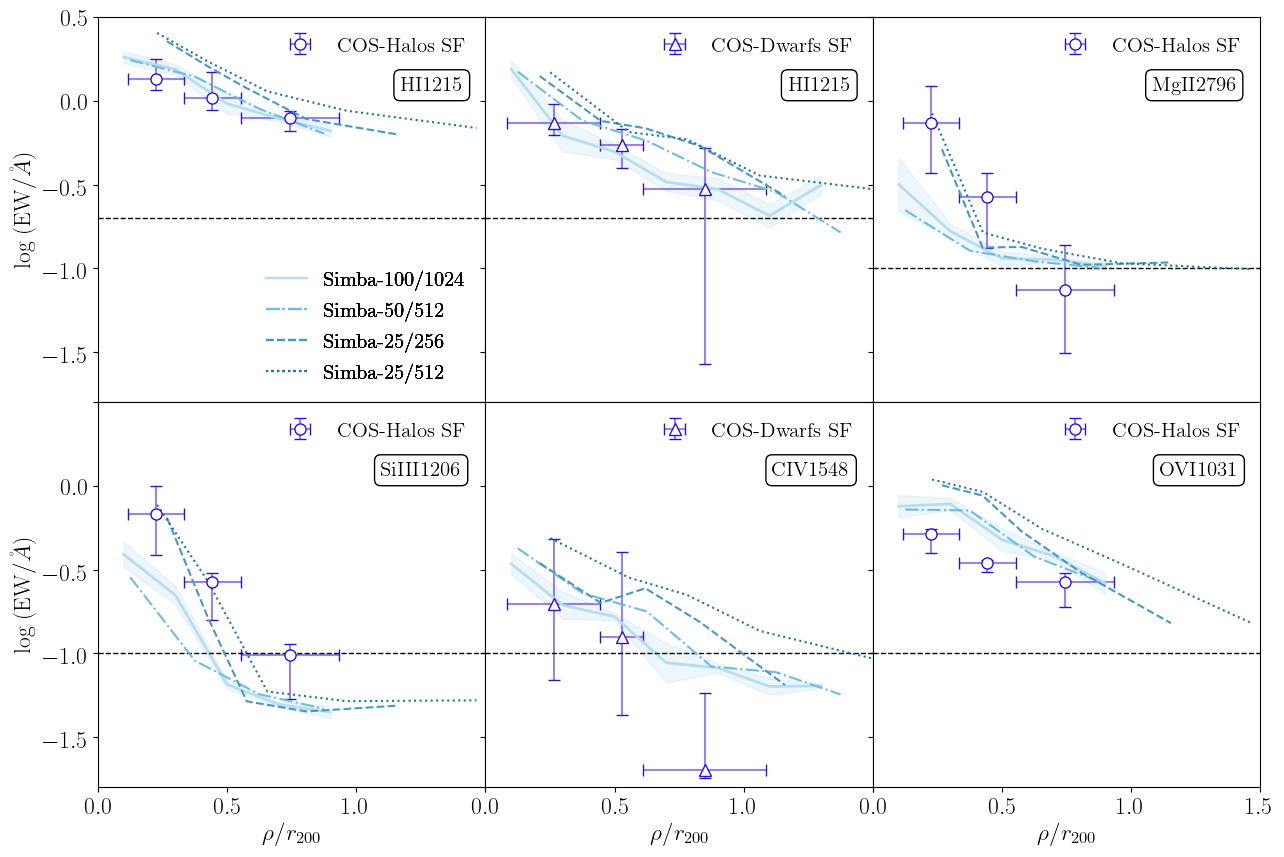}
	\vskip-0.1in
    \caption{Equivalent width of \HI and selected metal lines (as indicated in each panel) against $r_{200}$-scaled impact parameter for the star forming COS-Halos and COS-Dwarfs galaxy samples (circles and triangles, respectively) in different resolution \simba runs: fiducial 100 $\hmpc$ volume (solid); 50 $\hmpc$ volume with $512^3$ particles (dot-dashed); 25 $\hmpc$ volume with $256^3$ particles (dashed); higher resolution 25 $\hmpc$ volume with $512^3$ particles (dotted). Shaded regions show the cosmic variance uncertainties. Dark blue points represent star forming galaxies in the COS samples; vertical error bars represent the 25th and 75th percentiles of the data, while horizontal error bars indicate the width of the bins. Black horizontal dotted lines indicate the detection threshold of each line under the COS survey conditions. Both decreased volume and increased resolution separately lead to increased absorption in the CGM.} 
    \label{fig:resolution}
\end{figure*}

Figure \ref{fig:resolution} shows median equivalent width of selected ions against $r_{200}$-scaled impact parameter for the COS-Halos and COS-Dwarfs \simba galaxy samples using different resolution \simba runs: fiducial 100 $\hmpc$ volume (solid); 50 $\hmpc$ volume with $512^3$ particles (dot-dashed); 25 $\hmpc$ volume with $256^3$ particles (dotted); higher resolution 25 $\hmpc$ volume with $512^3$ particles (dashed). We assume an \citetalias{faucher-giguere_2020} ionising background. Light blue lines represent star forming galaxies in the \simba sample; shaded regions show the cosmic variance uncertainties. Dark blue points represent star forming galaxies in the COS samples; vertical error bars represent the 25th and 75th percentiles of the data, while horizontal error bars indicate the width of the bins. Black horizontal dotted lines indicate the detection threshold of each line under the COS survey conditions.

Comparing the 100 and 50 $\hmpc$ runs, the closeness of the results from the COS-Halos sample shows that these simulations are converged in terms of volume. Hence we are probing a large enough scale to capture the processes that suppress absorption in the CGM of \lstar galaxies. For the COS-Dwarfs sample there is some difference between the 100 and 50 $\hmpc$ runs; the lack of heating by the strongest AGN in the lower volume simulations leads to an increase in absorption around sub-\lstar star forming galaxies. Comparing the 50 and 25 $\hmpc$ runs, we see that absorption around star forming in the 25 $\hmpc$ volume with $256^3$ particles is increased in the 25 $\hmpc$ run, for all ion species and in both COS-Halos and COS-Dwarfs samples. Hence a larger, more representative volume leads to less absorption, indicating a lack of volume convergence in the lower volume simulations.

Meanwhile, increasing the resolution from $256^3$ to $512^3$ particles leads to a further increase in absorption, indicating that our simulations are not ideally converged, although the increase is typically $\la\times 2$ (at a given box size) which is comparable to other systematics in the comparison. 
\mycolor{In the COS-Halos sample, \HI and \OVI show a moderate absorption increase at high impact parameters due to resolution. However, at low impact parameters the increase due to resolution is small for galaxies in the COS-Halos sample, suggesting that these results are close to resolution convergence.}
Galaxies in the COS-Dwarfs sample see a particularly large increase in absorption across the entire range of impact parameters, perhaps since these low mass galaxies are more poorly resolved in the fiducial simulation. In summary, both reduced volume and increased resolution lead to significant changes in the absorption structure of the CGM, suggesting that both high resolution and sufficient volume to capture AGN heating sources are required to properly reproduce CGM statistics in simulations, which is numerically challenging.

% Don't change these lines
\bsp	% typesetting comment
\label{lastpage}
\end{document}